\documentclass[fleqn,usenatbib]{mnras}

\usepackage{newtxtext,newtxmath}

\usepackage[T1]{fontenc}

\DeclareRobustCommand{\VAN}[3]{#2}
\let\VANthebibliography\thebibliography
\def\thebibliography{\DeclareRobustCommand{\VAN}[3]{##3}\VANthebibliography}

\usepackage{graphicx}	
\usepackage{amsmath}	
\usepackage{orcidlink}

\defcitealias{Dekel2006}{DB06}
\defcitealias{Birnboim2003MNRAS.345..349B}{BD03}
\defcitealias{Stern2020}{SFFQ20}
\defcitealias{Stern2019}{SFFQ19}

\title[Early hot corona onset]{Early onset of the hot circumgalactic medium around Milky Way galaxies}

\author[Rufo-Pastor et al.]{
Víctor Rufo-Pastor\textsuperscript{\orcidlink{0000-0003-3259-9890}},$^{1}$\thanks{E-mail: victor.rufo$\_$pastor@fysik.lu.se}
 Oscar Agertz\textsuperscript{\orcidlink{0000-0002-4287-1088}},$^{1}$ Santi Roca-Fábrega\textsuperscript{\orcidlink{0000-0002-6299-152X}},$^{1}$ Martin P. Rey\textsuperscript{\orcidlink{0000-0002-1515-995X}},$^{2}$ Joss Bland-Hawthorn\textsuperscript{\orcidlink{0000-0001-7516-4016}}, $^{3}$
\newauthor Justin I. Read\textsuperscript{\orcidlink{0000-0002-1164-9302}}, $^{4}$ Thor Tepper-García\textsuperscript{\orcidlink{0000-0002-1081-883X}},$^{3}$ Corentin Cadiou\textsuperscript{\orcidlink{0000-0003-2285-0332}} $^{5}$ and Anatole Storck\textsuperscript{\orcidlink{0009-0000-3245-7951}} $^{6}$.\\ 
$^{1}$Lund Observatory, Division of Astrophysics, Department of Physics, Lund University, Box 118, SE-221 00 Lund, Sweden\\
$^{2}$University of Bath, Department of Physics, Claverton Down, Bath, BA2 7AY, UK \\
$^{3}$Sydney Institute for Astronomy, School of Physics, University of Sydney, Sydney, NSW 2006, Australia\\
$^{4}$University of Surrey, Guildford GU2 7XH, UK\\
$^{5}$Institut d'Astrophysique de Paris, Sorbonne Université, CNRS, UMR 7095, 98 bis bd Arago, 75014 Paris, France\\
$^{6}$Sub-department of Astrophysics, University of Oxford, DWB, Keble Road, Oxford OX1 3RH, UK 
}

\date{Accepted XXX. Received YYY; in original form ZZZ}

\pubyear{2026}

\begin{document}
\label{firstpage}
\pagerange{\pageref{firstpage}--\pageref{lastpage}}
\maketitle

\begin{abstract}
Hot coronae are key regulators in Milky Way (MW)-type galaxy formation and are thought to emerge once a galaxy’s dark matter halo crosses a critical mass threshold, $M_{\rm vir}\sim10^{12}\,{\rm M}_\odot$, so that accretion shocks can heat infalling gas to virial temperatures, $T_{\rm vir}$. Here we investigate the roles of stellar feedback and accretion shocks in establishing the circumgalactic medium (CGM) around MW-mass galaxies using the {\small VINTERGATAN} cosmological simulation suite. We find that stellar feedback drives the early formation of a hot CGM, with $T_{\rm CGM}\sim T_{\rm vir}$. By contrast, in feedback-free simulations, accretion shocks alone do not produce a CGM with comparably high average temperatures at early epochs ($z>2$), independent of gas metallicity. We show that CGM virialisation (defined when halo-gas cooling time exceeds the free-fall time) begins with the emergence of a hot corona ($T>10^5$ K) that can appear as early as $z\approx3$ at virial radius, $R_{\rm vir}$, when feedback is included. This is much earlier than predicted by accretion-shock theory or feedback-free models ($z\lesssim1$). While accretion shocks alone produce an inside-out virialisation pattern, feedback can alter this behaviour, driving a rapid transition ($\lesssim0.5$ Gyr) that is either outside-in or nearly simultaneous throughout the halo. Finally, these early-forming coronae are largely stable from formation through to the present day, but can appear to evolve simply because the definition of $R_{\rm vir}$ grows with time. This ‘CGM pseudo-evolution’ is analogous to the pseudo-evolution of dark matter haloes and should be taken into account when interpreting the evolution of hot coronae.

\end{abstract}




\section{Introduction}\label{sec:intro}

The warm-hot gas ($T > 10^5$ K) that fills the circumgalactic medium (CGM) of galaxies such as the Milky Way (MW) is commonly referred to as the galactic hot corona. The existence of diffuse hot gas around galaxies was first proposed by \citet[][]{Spitzer1956} \citep[see also ][]{Bahcall69} and has since become a central ingredient of modern galaxy formation theory \citep{Rees1977MNRAS.179..541R, WhiteRees1978MNRAS.183..341W, Fall1980MNRAS.193..189F}. The CGM acts as the interface between galaxies and the intergalactic medium (IGM), where gas accreted from the cosmic web interacts and mixes with material expelled by galactic feedback processes \citep{Lilly2013ApJ...772..119L, Dekel2014MNRAS.444.2071D, Somerville2015ARA&A..53...51S, Ayromlou2023MNRAS.524.5391A}. This continuous exchange regulates star formation and drives the baryon cycle of galaxies \citep{Tumlinson17review, Wright2024}. Understanding when and how a hot corona forms, and how this modifies the mode through which galaxies acquire gas and regulate star formation \citep{Kere2005,Brooks2009ApJ...694..396B, Stern2021, Sultan2026}, is therefore essential for understanding galaxy growth.

The formation of hot coronae is commonly discussed in the context of the transition from cold- to hot-mode accretion \citep{Katz2003ASSL..281..185K,Birnboim2003MNRAS.345..349B, Kere2005,Dekel2006}. This transition depends on the competition between radiative cooling, gravity, and pressure support. A useful way to characterize this balance is through the cooling time, $\tau_\text{cool}$, which measures how long gas requires to radiate away its thermal energy, and the free-fall time, $\tau_\text{ff}$. When cooling is inefficient ($\tau_\text{cool}>\tau_\text{ff}$), shocked gas can remain pressure-supported and develop into a hot atmosphere. Conversely, when cooling proceeds faster than gravitational collapse ($\tau_\text{cool}<\tau_\text{ff}$), the thermal energy generated by shocks is rapidly lost and gas accretes predominantly in a cool phase.

This physical picture was formalized by \citet{Birnboim2003MNRAS.345..349B} and \citet{Dekel2006}, hereafter \citetalias{Birnboim2003MNRAS.345..349B} and \citetalias{Dekel2006}, through the virial-shock framework. In this model, dark matter haloes above a characteristic mass of $\sim 10^{11.5}-10^{12}~\text{M}_\odot$ can sustain stable accretion shocks that heat infalling gas to approximately the virial temperature, producing a volume-filling hot corona. The predicted transition mass is only weakly dependent on redshift and is broadly consistent with the characteristic halo masses at which cosmological simulations \citep{Kere2005,Kere2012, Voort2011, Claude2011, Stern2021} and idealized simulations \citep{Fielding2017, Stern2020} find a shift from predominantly cold to predominantly hot accretion.

Unlike the analytical framework, however, cosmological simulations follow the fully 3D evolution of gas and include additional physical processes such as stellar and/or AGN feedback, anisotropic accretion, and mergers. Feedback can strongly modify the thermal structure of the CGM; the original \citetalias{Dekel2006} model assumes 1D spherically symmetric evolution and neglects these processes. Despite these differences, simulations often recover a transition mass remarkably similar to that predicted by the virial-shock model. Whether this agreement indicates that virial shocks set the fundamental scale for hot-corona formation, with feedback mainly regulating the resulting CGM structure, or whether different physical mechanisms conspire to produce a similar transition mass, remains an open question. 

Addressing this question requires distinguishing the relative contributions of the two main mechanisms capable of heating the CGM of Milky Way-mass haloes: gravitational accretion shocks and feedback-driven outflows. Both processes can provide pressure support and increase the thermal energy of circumgalactic gas. 
In particular, observations and simulations show that star formation-driven winds can reach velocities comparable to those of cosmological inflows \citep{Veilleux2005,
Voort2012MNRAS.423.2991V,Chisholm2015ApJ...811..149C,Goerdt2015MNRAS.450.3359G,Muratov2017MNRAS.468.4170M,Fox2019ApJ...884...53F},
implying that feedback-driven shocks may be comparable in strength to those generated by gravitational accretion. Consequently, identifying the dominant mechanism responsible for establishing a hot CGM remains challenging.
The coexistence of these two heating channels complicates the interpretation of the apparent agreement between cosmological simulations and the \citetalias{Birnboim2003MNRAS.345..349B,Dekel2006} framework. 

Numerous studies find that Milky Way-mass haloes transition from predominantly cold to predominantly hot accretion at halo masses of a few $10^{11}-\!10^{12}\ {\rm M}_\odot$, broadly consistent with the virial-shock predictions, despite substantial differences in numerical methods and feedback prescriptions
\citep{Ocvirk2008MNRAS.390.1326O, Khalatyan2008MNRAS.387...13K, Brooks2009ApJ...694..396B,Kere2009MNRAS.395..160K,Kere2009}. However, the redshift at which this transition occurs for $z=0$ MW-mass haloes varies considerably, typically spanning $z\approx1\!-\!3$, with some studies finding substantial hot gas fractions already at $z\approx4$. Moreover, cold and hot accretion modes often coexist over extended periods, making it difficult to define a unique epoch for hot-corona formation. The predicted transition is also sensitive to how gas phases are identified, as temperature, density, and metallicity cuts can lead to different interpretations of the thermal state of the CGM, and these properties depend on the adopted feedback implementation \citep{StrawnAGORA2024ApJ...962...29S}.

The impact of feedback further complicates this picture by altering the structure and thermal history of the CGM. Stronger stellar feedback can increase cooling times and hot-gas fractions, potentially producing a volume-filling hot atmosphere earlier than expected from gravitational heating alone \citep[e.g.][]{Correa2018MNRAS.473..538C}. Conversely, feedback-driven winds may initially heat the outer halo before cooling times in the inner CGM become sufficiently long to sustain a fully virialized atmosphere \citep[Sec. 2.1.3]{Faucher2023ARA&A..61..131F}. Therefore, although simulations broadly reproduce a characteristic transition mass similar to that predicted by the virial-shock framework, they do not yet establish whether gravitational heating or feedback is primarily responsible for creating the hot CGM.

A related, but distinct, question concerns the virialisation of the CGM. While the previous discussion focuses on the physical mechanisms responsible for establishing a hot corona, an equally important issue is whether the presence of hot gas necessarily implies the onset of hydrostatic support. Work based on cooling-flow arguments \citep{Mathews1978ApJ...224..308M,Fabian1984Natur.310..733F,Bertschinger1989ApJ...340..666B, Stern2019} have emphasized the ratio $\tau_\text{cool}/\tau_\text{ff}$ as a diagnostic of whether the CGM can maintain approximate hydrostatic equilibrium \citep{Stern2020, Stern2021}. Although the characteristic halo masses associated with virialisation are similar to those predicted by virial-shock arguments, these two transitions do not necessarily represent the same physical process.

A halo may contain a substantial hot gas component while still experiencing significant cold accretion, or may satisfy the conditions for hydrostatic support in specific regions while remaining dynamically complex elsewhere. Conversely, the establishment of a hot atmosphere does not automatically imply that the CGM has reached a fully virialised state. Whether hot-corona formation and CGM virialisation tends to occur simultaneously, or instead represent distinct stages in the thermal evolution of galactic haloes, therefore remains an open question.

Motivated by these open issues, in this work we investigate when and how hot coronae emerge in Milky Way-mass haloes and how this process relates to CGM virialisation using a suite of cosmological simulations specifically designed to isolate the effects of stellar feedback, gas metallicity, and halo assembly history. By varying these ingredients independently, we assess the relative importance of gravitational accretion and stellar feedback in establishing the hot CGM, evaluate the predictive power of the virial-shock framework, and determine whether hot-corona formation and CGM virialisation occur simultaneously or represent distinct stages in the thermal evolution of the halo.

The article is structured as follows. In Sec.~\ref{sec:method}, we describe the simulation suite, the virial-shock model, and the definitions of the physical quantities used throughout this work. In Sec.~\ref{sec:results}, we examine the physical mechanisms responsible for establishing a hot CGM and assess the predictive power of the virial-shock framework. We further explore whether hot-corona formation and CGM virialisation occur simultaneously or represent distinct stages in the thermal evolution of the halo. We discuss the implications of these findings in Sec.~\ref{sec:DISCUSSION} and summarize our main conclusions in Sec.~\ref{sec:summary}.


\section{Methodology} \label{sec:method}
\subsection{Simulation models} \label{sec:method_sim}
\begin{figure}
 \centering
 \includegraphics[width=0.45\textwidth]{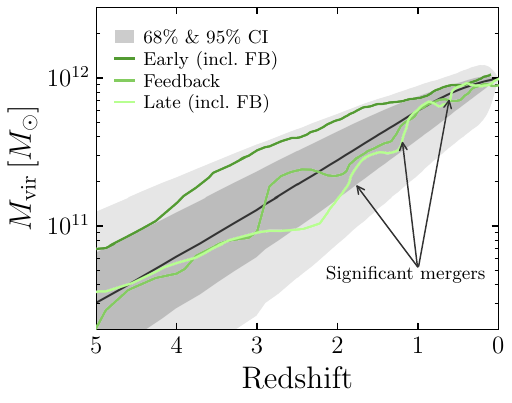}
    \caption{Growth of $M_{\rm vir}$ as a function of redshift for the simulations from the {\small VINTERGATAN} suite used in this study. Solid curves show the fiducial FB runs. Following \protect\cite{Martin2022MNRAS.510.4208R}, we compare these histories with the median mass-growth history of 28475 MW-mass haloes selected from IllustrisTNG (solid black line). Data provided by Martin Rey (\textit{private communication}, see main text). The dark and light grey shaded regions show the 68\% and 95\% confident intervals (CI) of the TNG population.}
    \label{FIG:MvirEvol}
\end{figure}

In this work, we study six zoom-in cosmological hydrodynamical simulations of Milky Way--mass haloes from the {\small VINTERGATAN} project \citep[specifically the {\small VINTERGATAN-GM} suite from][following the original {\small VINTERGATAN} simulation described in \citealt{Agertz2021,Renaud2021,Renaud2021MNRAS.503.5868R,Segovia2022MNRAS.516.2272S}]{Martin2022MNRAS.510.4208R,Rey2023}, carried out with the adaptive mesh refinement code {\small RAMSES} \citep[][]{Teyssier2002A&A...385..337T}. We consider three fiducial simulations, corresponding to three different halo assembly histories, classified as early-forming ({\it Early}), intermediate-forming ({\it Feedback}; for the sake of clarity), and late-forming ({\it Late}). The Early halo assembles rapidly and experiences its main merger at high redshift ($z>3$), the Feedback halo undergoes a significant merger at an intermediate epoch ($2<z<3$), and the Late halo has the most delayed assembly history, with significant merger events occurring at comparatively low redshift ($z<1.5$). The initial conditions of the Early and Late runs are described in \citep{Martin2022MNRAS.510.4208R}, and the Feedback run in \citep{Rey2023}. For clarity, the influence of different halo assembly histories is discussed separately in Appendix~\ref{sec:ap_AMH}.
We consider three additional simulations based on the initial conditions and halo evolution of the Feedback run. In each of these, stellar feedback is deactivated and the gas metallicity fixed: {\it NoFB$10^{-3}Z_\odot$}, {\it  NoFB$0.1Z_\odot$}, and {\it NoFB$1Z_\odot$}, corresponding to metallicities of $10^{-3}Z_\odot$, $10^{-1}Z_\odot$, and $Z_\odot$, respectively. This allows us to remove the impact of stellar feedback and isolate the influence of accretion shocks. The virial mass growth histories of all haloes are shown in Fig.~\ref{FIG:MvirEvol}, together with a sample of 28475 Milky Way--mass host dark matter haloes from the IllustrisTNG simulation suite (\citealt{Nelson2018MNRAS.475..624N}; see \citealt{Martin2022MNRAS.510.4208R} for details). Note that we did not include in Fig.~\ref{FIG:MvirEvol} the NoFB runs as their evolutionary path is the same as that of the Feedback run. 

The {\small VINTERGATAN} simulations achieve a maximum spatial resolution of $\sim 20$ pc, in the dense interstellar medium (reaching $\rho \gg 1\ {\rm cm}^{-3}$ and $T<100 {\rm\ K}$). The Lagrangian region is selected from a 50$h^{-1}$ Mpc (or $\sim 73$\,Mpc for our choice of $h$, see below) cosmological box to include a MW-type halo. The cosmological model assumed is $\Lambda$CDM with cosmological parameters $\Omega_m=0.3139$, $h=0.6727$, $\sigma_8=0.8440$, $n_s=0.9645$ from \cite{PLANCK2016A&A...594A..13P}. The dark matter mass resolution in the zoom-in regions is $m_{\text{DM}}= 2\times 10^5 {\rm M_\odot}$ and $m_\star = 10^4 {\rm M_\odot}$ for stellar particles at formation. Star formation is modelled as a Poisson process where star particles are generated based on a cell-by-cell method with a star formation rate density described by $\dot{\rho}_\star = \epsilon_{\text{ff}}\dfrac{\rho_{\text{gas}}}{t_\text{ff}}$ if the gas density in the cell satisfies $\rho_{\text{gas}}>100\ \text{cm}^{-3}$ and the gas temperature, $T_{\text{gas}}<500$~K. Here $t_{\text{ff}}=\sqrt{3\pi/32G\rho_{\rm gas}}$ is the local free-fall time, and $\epsilon_\text{ff}$ is the local star formation efficiency per free-fall time, set to 10 percent \citep[see][]{Grisdale2017, Grisdale2018, Agertz2020}. The implemented stellar feedback processes include stellar winds, radiation pressure, type II, and type Ia supernovae \citep{AgertzKravtsov2013ApJ...770...25A,AgertzKravtsov2015ApJ...804...18A, AgertzKravtsov2016ApJ...824...79A}. The SNe feedback is injected as thermal energy if there is sufficient grid resolution for the SN cooling radius to be resolved. If not, the explosion is initialized in momentum conserving phase (see \citealt{Kim2015ApJ...802...99K, Martizzi2015MNRAS.450..504M, Agertz2021}). Additionally, heating due to the ultraviolet background radiation is modelled using an updated version of \citealt{Haardt1996ApJ...461...20H}, as implemented in the public version of {\small{RAMSES}}, assuming a reionisation epoch at $z=8.5$, see \citet{Rey2023} for details. The gas cooling is metallicity-dependent. For temperatures  $T>10^4$K, the cooling follows \cite{SutherlandDopita1993ApJS...88..253S}, while for $T<10^4$K the rates of \cite{Rosen1995ApJ...440..634R} are applied.

We identify the halo centres using the halo finder described in \citep{RAMON2025A&A...698A.303R} that combines {\small ROCKSTAR} with Consistent Trees and kinematics corrections. We ensure we are tracking the centre of the host galaxy using the ``shrinking sphere'' method \citep{Power2003MNRAS.338...14P} and tracking the 100 most gravitationally bound star particles. Once the centre is established, we compute the virial radius, $R_\text{vir}$, and the virial mass $M_\text{vir}$ (defined here as the total DM mass enclosed within a sphere of radius $R_\text{vir}$) based on the definition provided by \cite{Bryan1998ApJ...495...80B}. We use the same halo finder to identify satellites and remove gas within $0.2R_{\rm vir}$ of their centres, choosing this radius as a conservative proxy for enclosing the ISM region. For the main DM halo we compute its scale radius, $r_{\rm s} = R_{\rm vir}/c$, where the concentration parameter $c$ is computed following \cite{Wang2024MNRAS.52710760W}, whose method is less sensitive to limited particle resolution than conventional NFW-profile fitting.

We define the CGM as the volume of a spherical shell centred on the host halo with radii between 0.2$R_\text{vir}$$\ $and $R_\text{vir}$, as is commonly done in the literature \cite{Ocvirk2008MNRAS.390.1326O}\footnote{However, the inner CGM definition differs between studies, e.g. 0.1$R_\text{vir}$ \citep{Stern2021, Sultan2024arXiv241016359S}.}. This geometrical criterion for the inner CGM allows us to analyse the disc vicinity while avoiding the overlap with the ISM across time. Within this volume, the minimum and median spatial resolution are $\sim 20$ and $\sim 1100~\mathrm{pc}$, respectively.

In this study, we make a key distinction between the hot corona and the virialized CGM. We define the hot corona as the extended warm-hot gas with $T\!>\!10^5$ K \citep[similar to other works, e.g.,][]{Correa2018MNRAS.473..538C,Hopkins2023,Wright2024,Sultan2024arXiv241016359S, Sankar2025arXiv251103793S}, and generally refer to gas with $T<10^5 {\rm K}$ as cold. Here, we use the term “thermalised” to denote the sustained presence of this hot phase in the CGM, and hence the thermal state of the CGM, rather than the thermal history of individual gas elements. In contrast, we define CGM virialisation through the thermodynamic condition $\tau_{\rm cool}/\tau_{\rm ff} > 1$, following \cite{WhiteRees1978MNRAS.183..341W,Stern2020,Stern2021}.

\subsection{Cooling time and free fall time}\label{SEC:Method_tcooltff}
In this study, we want to study the hot corona onset and its connection to the virialisation of the CGM gas. The virialisation relates to the thermo-dynamical state
of the gas and it is defined as the ratio between the cooling time and the free fall time fulfilling, $\tau_\text{cool} / \tau_\text{ff}\!>\!1$, \citep[][]{Rees1977MNRAS.179..541R, WhiteRees1978MNRAS.183..341W, Hopkins2023}. If $\tau_\text{cool}\!>\!\tau_\text{ff}$, the volume-filling hot phase is able to maintain a gas temperature $T\!\sim\!T_{\rm vir}$, where 

\begin{equation}
T_\text{vir} = \frac{\mu m_{\rm p}G^{2/3}\Delta_{\rm c}^{1/3}\Omega^{1/3}H_0^{2/3}}{16^{1/3}k_{\rm B} }M_{\rm vir}^{2/3}(1+z)  \  .
\label{eq:tvir}
\end{equation}
Here $m_{\rm p}$ is the proton mass, $\mu$ is the mean molecular weight of the gas, $G$ is the gravitational constant, $\Omega$ is the matter density parameter, $H_0$ is the Hubble constant at redshift zero, $k_{\rm B}$ is the Boltzmann constant, and the overdensity $\Delta_c$ is calculated following \cite{Bryan1998ApJ...495...80B} in a universe with $\Omega_{\rm R}\! =\!0;\ \Omega_{\rm m}+\Omega_\Lambda\! =\!1$. The cooling time is computed as
\begin{equation}
    \tau_{\text{cool}} = \dfrac{3}{2}\dfrac{p}{n_H^2  \Lambda_\text{net}(T,Z)}
\label{eq:tcool}
\end{equation}
where $p$ is the pressure, $n_H$  is the number density of hydrogen, and $\Lambda(T,Z)$ is the temperature-metal-dependent cooling function (see Sec. \ref{sec:method_sim}). 

For the free-fall time we adopt  
\begin{equation}
    \tau_{\text{ff}} = \sqrt{2}\dfrac{r}{v_{\text{cir}}},
    \label{eq:tff}
\end{equation}
where $r$ is the radial distance from the galactic centre, and $v_{\text{cir}} = \sqrt{G M(<r)/r}$ is the circular velocity at radius r, and $M(<r)$ the enclosed mass. This quantity represents the characteristic time for gas at radius $ r$ to fall toward the centre in the gravitational potential of the halo. Here, we assume a \emph{constant} gravitational acceleration following \citealt{McCourt2012MNRAS.419.3319M, Stern2021}. 

All of the above-mentioned quantities entering $\tau_{\rm cool}$ and $\tau_{\rm ff}$ are extracted for each gas cell, and the median $\tau_{\rm cool}/\tau_{\rm ff}$ is then computed within radial shells, weighted by gas mass or volume.
\begin{figure*}
    \centering
    \includegraphics[width=0.9\textwidth]{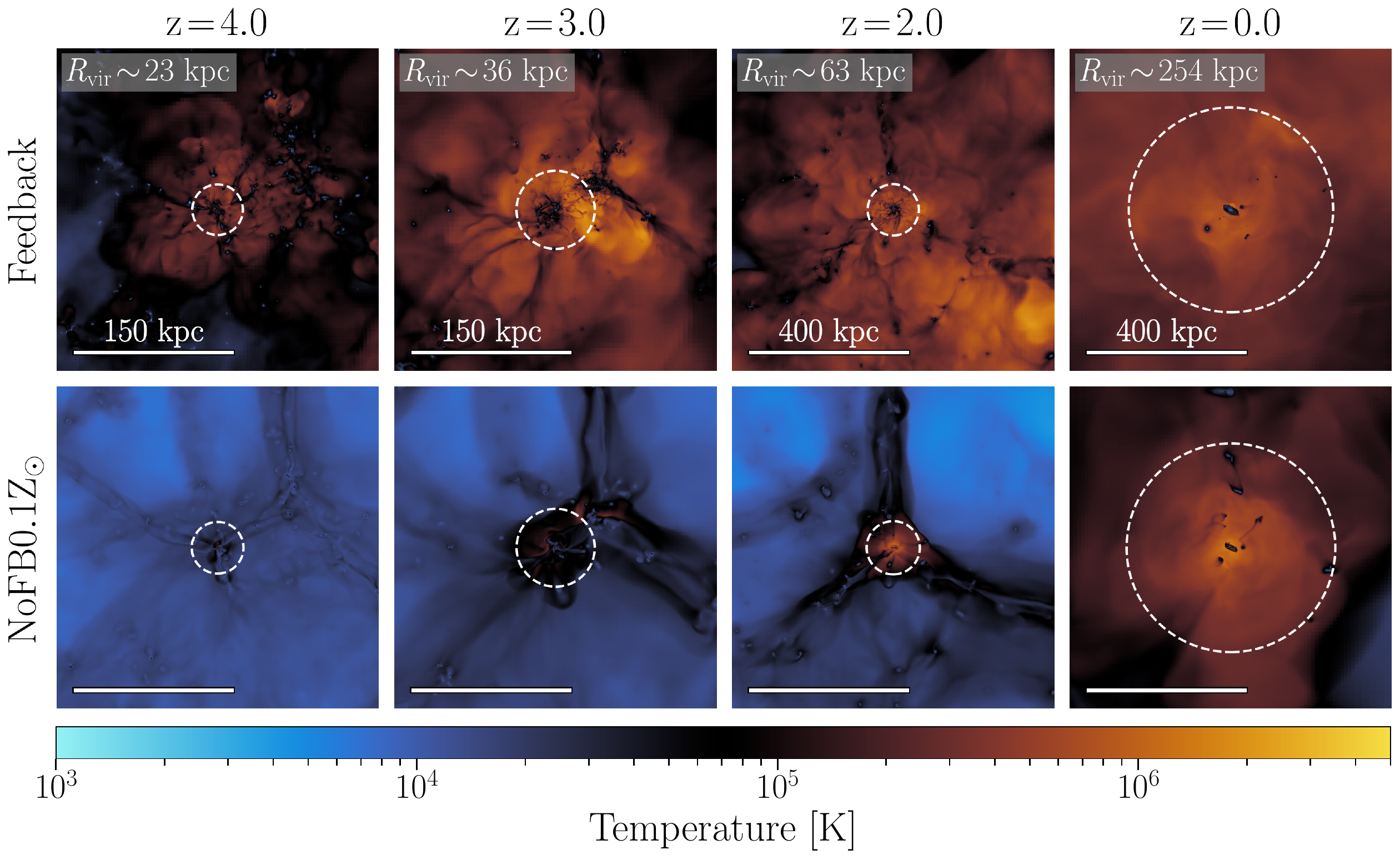}
    \caption{
     Density-projected gas coloured according to temperature at four selected cosmological epochs, $z=4,\ 3,\ 2,$ and $0$ (from left to right). The corresponding value of the temperature is indicated by the colour scale at the bottom. Each row corresponds to a specific simulation as indicated by the label to the left of each row. The $R_\text{vir}$ is indicated by a white dashed circle in each panel. It can be seen in the top row that stellar feedback quickly heats the gas well beyond $R_\text{vir}$, while accretion shocks alone (bottom row) build up the hot corona at later times and are highly sensitive to the metal content in the CGM.}
    \label{FIG:MULTITEMP}
\end{figure*}

\subsection{The Virial Shock model} \label{sec:virshock}
We use the Virial Shock model by \citetalias{Birnboim2003MNRAS.345..349B} and \citetalias{Dekel2006} to obtain a benchmark for when a hot corona is expected to develop via pure accretion shocks, i.e. without stellar feedback, in our simulations. In this way we can contrast the analytical result of this 1D model with the evolution from the simulations both with and without feedback. In practice, this corresponds to analysing whether our haloes reach the condition $M_\text{vir}\! >\! M_{\rm shock}$ (see Eq. \ref{eq:Mshock}), for which $\tau_\text{cool}/\tau_\text{ff}\!\gg\!1$ is met and a thermally stable, hot corona can be sustained. We therefore compute the analytical $M_{\rm shock}$ using the haloes' properties (metallicities and the baryon fraction) from each of our simulations, which we label $M_{\rm shock,VG}$. 

Our calculation is based on the shock-stability criterion $\tau_\text{cool}=\tau_\text{comp}$, where $\tau_\text{comp}$ is the compression time of the gas in the post-shock region, defined as $\tau_{\rm comp} = (21\ \rho)/(5\ \dot{\rho})$ (eq. 2 in \citetalias{Dekel2006}). This criterion leads to the critical post-shock temperature\footnote{We report a typo in eq. 6 of \citetalias{Birnboim2003MNRAS.345..349B}: the correct expression for the post-shock temperature is $T_1=\frac{\mu}{k_\text{b}N_\text{a}}\frac{2(\gamma-1)}{(\gamma+1)^2}u_0^2$, rather than $T_1=\frac{\mu}{k_\text{b}N_\text{a}}\frac{2\gamma-1}{(\gamma+1)^2}u_0^2$.} required for a stable accretion shock at radius $r$ \citepalias[eqs. 19 and 24]{Dekel2006}, or a stable virial shock at $R_\text{vir}$,

\begin{equation}
    T_{6,\text{crit}} = 1.04\ A(z)^{-3/2}\ F(M,z)\ \Lambda_{-22}(T,Z,z) \,
    \label{eq:Tcrit}
\end{equation}
where $T_6 = T / 10^6\,{\rm K}$, 
\begin{equation}
    A(z) \equiv \frac{1}{1+z}
    \left[
        \frac{\Delta_{\rm c}}{200}
        \frac{\Omega_{\rm m}}{0.3}
        \left(\frac{h}{0.7}\right)^2
    \right]^{-1/3},
    \label{eq:Az}
\end{equation}
and $F(M,z)$ is a dimensionless halo structure function \citepalias[eqs. 30 and A7 in][for further details]{Dekel2006}, and $\Lambda_{-22}$ is the cooling function in units of $10^{-22}\,{\rm erg\,cm^3\,s^{-1}}$ that depends on gas temperature $T$, the gas metallicity $Z$, and the cosmological epoch $z$. The factor $F$ collects several parameters describing halo properties, e.g. the concentration parameter of the DM profile and the baryon fraction \citepalias[see eq. 30 in][]{Dekel2006}. To keep our calculation of $M_\text{shock}$ consistent with \citetalias{Dekel2006} we adopt the cooling tables from \cite{SutherlandDopita1993ApJS...88..253S}, although these do not include Compton scattering of the cosmic microwave background or the effect of external radiation on the cooling rate via hydrogen reionisation.
 
To go from $T_{6,\text{crit}}$ to $M_\text{shock}$, firstly we use eq. 15 of \citetalias{Dekel2006} that connects the post-shock temperature with the pre-shock radial velocity; the so-called Rankine-Hugoniot shock jump conditions (assuming monoatomic gas, and using the index 0, 1 for pre-shock and post-shock gas, respectively),
\begin{equation}
\frac{k T_1}{m} = \frac{3}{16} \, u_0^2 \left(1 + \tilde{u}_s \right)^2
\label{eq:15}
\end{equation}
so $T_1$ is the shocked gas temperature, $u_0$ is the radial velocity of the infalling gas, and $\tilde{u}_s = u_s/|u_0|$ is the ratio between the radial shock velocity and $u_0$. Then we assume a \textit{homologous} condition for the velocity field \citepalias[section 2.2 in][]{Birnboim2003MNRAS.345..349B}, so we can include the virial velocity via $u_0=f_\text{u}V_\text{vir}$ where $f_\text{u} = \frac{u_0(r)}{u_0(R_\text{vir})}$ \citepalias[eq. 27 in][]{Dekel2006}; which is essentially a ratio between velocities at radius $r$. Note that \citetalias{Dekel2006} uses $|u_0|=V_\text{vir}$ for crude estimations. 

To keep consistency with \citetalias{Dekel2006}, we assume a slowly moving shock, $\tilde{u}_s\rightarrow0$. This limit is appropriate for evaluating the onset of virial-shock stability because smaller shock velocities favour stabilization, whereas a rapidly expanding shock makes compression inefficient and stability harder to achieve. We also note that in \citetalias{Dekel2006}, the value of $\tilde{u}_s$ can shift the $M_\text{shock}$ up to 24\% when changing it from\footnote{The value $\tilde{u}_s=1/7$ is obtained using the jump condition for the radial velocity condition \citetalias[eq. 8 in][]{Dekel2006} assuming $\gamma = 5/3$.} 1/7 to 0, introducing a corresponding systematic uncertainty in our analytical results. We obtain
\begin{equation}
V_{\rm vir} = \left(\frac{16}{3}\frac{k_{\rm B}}{\mu m_{\rm p}}\frac{T_{\rm crit}}{f_u^2}\right)^{1/2},
\label{eq:Vvir}
\end{equation}
and using the virial relations we get \citepalias[eq. A6 in][]{Dekel2006}
\begin{equation}
M_{\rm shock} = 6.1 \times 10^{11}\,M_\odot \, A(z)^{3/2}
\left(\frac{V_{\rm vir}}{100\,{\rm km\,s^{-1}}}\right)^3 .
\label{eq:Mshock}
\end{equation}

For each cosmological epoch between $z=0$ and 5 we use a range of halo masses $M$ from $10^{10}-10^{13} \ {\rm M}_\odot$ and impose the shock stability condition to solve for $T_{\rm6, crit}$. The solution $M_{\rm shock}$ represents the halo mass at which a stable accretion shock can first be supported; then it sets how much the shock can expand in the CGM. Hence for a fixed radial distance, haloes with $M \!<\! M_{\rm shock}$ cool efficiently and cannot maintain a stable shock. Haloes with $M \!>\! M_{\rm shock}$ have cooling times longer than the compression time and thus sustain a stable shock.

 \begin{figure}
    \centering
\includegraphics[width=0.5\textwidth]{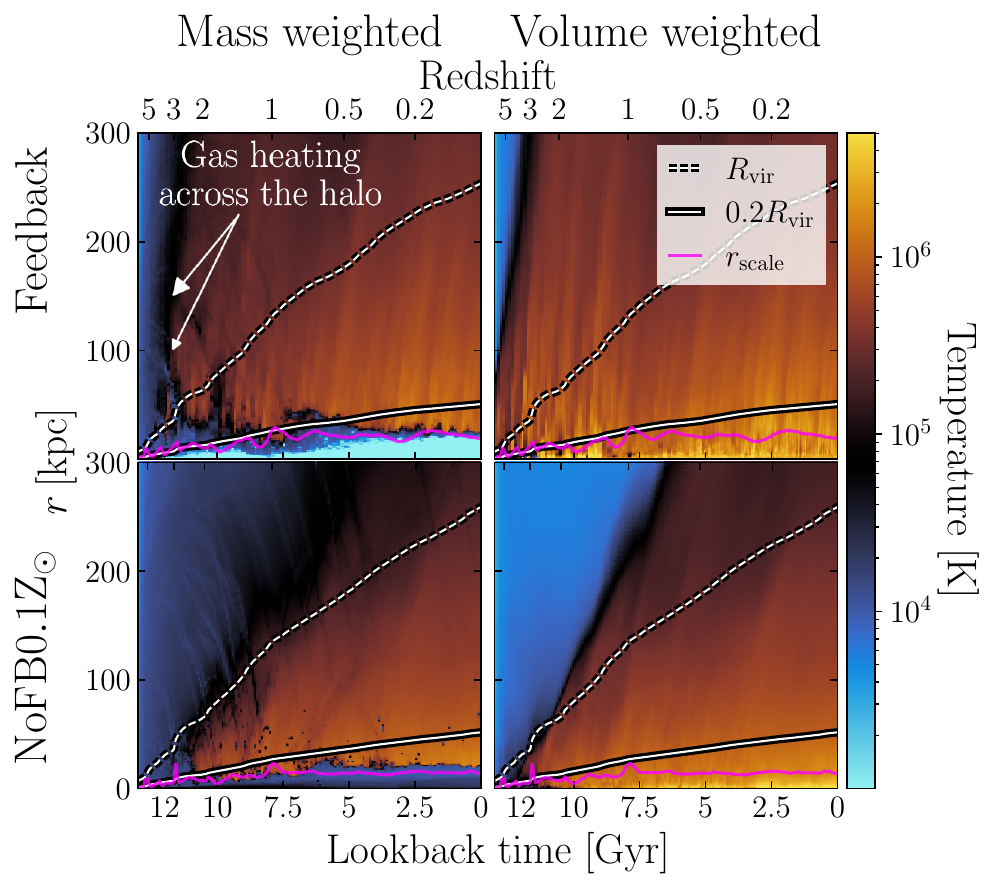}
    \caption{Left panels: radial distribution within 300 kpc (proper distance) from the halo's centre over time of the mass-weighted temperature for the Feedback run (top panel) and the NoFB0.1$Z_\odot$ run (bottom panel). Each panel presents the evolution of $R_{\text{vir}}$ (dotted white line) and $0.2R_{\text{vir}}$ (solid white line) to track the CGM. We also include the evolution of the DM halo's scale radius (magenta line) to track changes in the characteristic radial scale of the halo density profile. Right panels: same as left but volume weighted.}
    \label{FIG:HISTOGRAMS}
\end{figure}

\begin{figure*}
    \centering
  \includegraphics[width=0.9\textwidth]{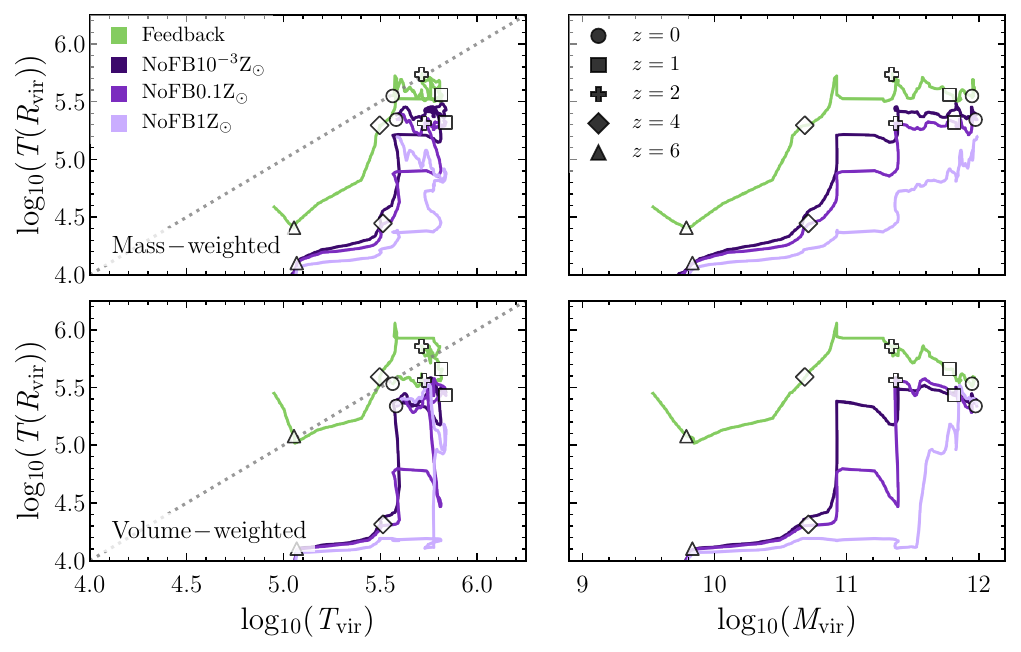}
    \caption{Top row:
    Gas mass-weighted median temperature in a shell centred at $R_{\text{vir}}$ relative to its value expected from shock heating $T_\text{vir}$ (left panel), and relative to the growth of $M_\text{vir}$ for the Feedback and NoFB runs. In each panel, the evolution in the Feedback run and in the NoFB run is indicated by green curve and a purple curve, respectively. We have included markers flagging our selected cosmological epochs and a dotted line marking a 1:1 relation in the left panels.}
    \label{FIG:TVIR_TatRVIR}
\end{figure*}

\begin{figure*}
    \centering
    \includegraphics[width=0.9\textwidth]{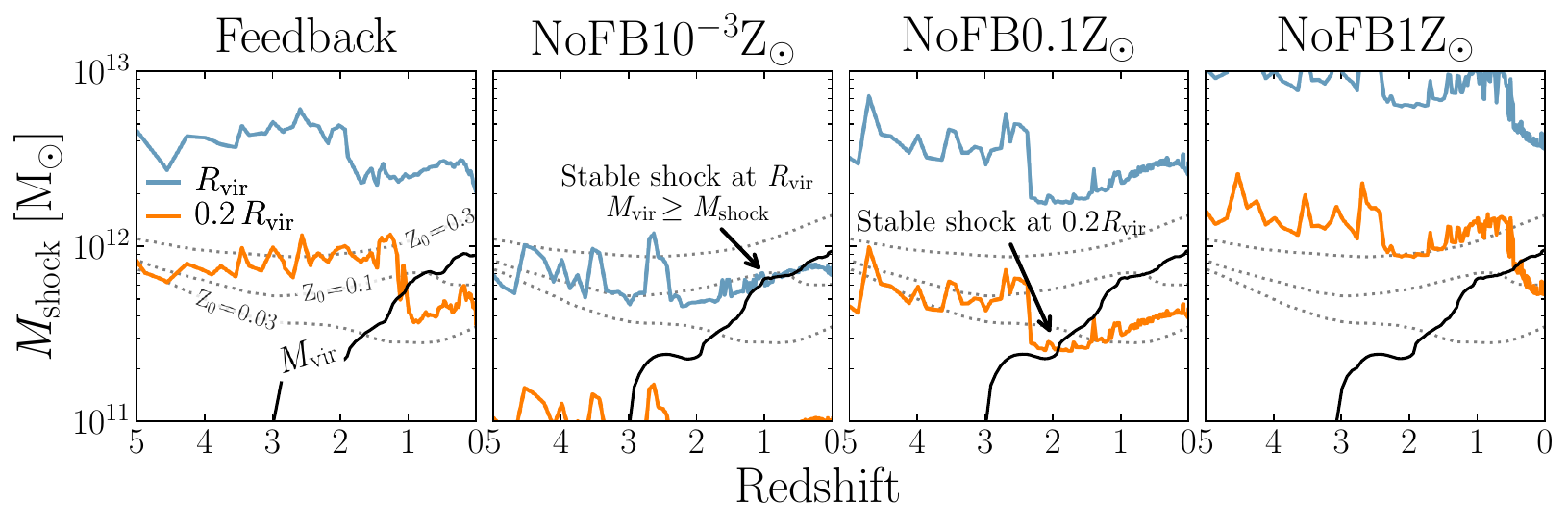}
    \caption{
     The expected virial mass required for a stable accretion shock, $M_\text{shock}$, is shown as a function of redshift for two CGM shells in the Feedback and NoFB runs: $R_{\text{vir}}$ (blue), marking the outer CGM, and $0.2R_{\text{vir}}$ (orange), tracing the inner CGM. The black solid line shows the virial-mass evolution, $M_\text{vir}$, in each run. A stable accretion shock is expected when $M_\text{vir}\!>\!M_\text{shock}$ at a given shell; therefore, the intersections between the black and coloured curves mark the predicted transition to accretion-shock heating. The arrows in the second and third panels indicate two examples of such transition points, where $M_\text{vir}=M_\text{shock}$: at $R_{\rm vir}$ at $z=1$ for NoFB10$^{-3}Z_\odot$, and at $0.2R_{\rm vir}$ at $z=2$ for NoFB0.1$Z_\odot$. Across all runs, the accretion-shock model shifts the predicted formation of the hot corona towards higher halo masses compared to the emergence of the hot CGM measured in the simulations (Figs. \ref{FIG:HISTOGRAMS} and \ref{FIG:TVIR_TatRVIR}), showing that accretion shocks alone cannot account for its earlier formation. For comparison with Fig.~2 of \protect\citetalias{Dekel2006}, the faint dotted lines show our calculation for a $0.1R_{\text{vir}}$ shell using fixed present-day metallicities of $0.3$, $0.1$, and $0.03\,Z_\odot$. These reference curves show how the classical prediction at $0.1R_{\text{vir}}$ compares with our time-dependent estimates at $0.2R_{\text{vir}}$ and $R_{\text{vir}}$.}
    \label{FIG:MSHOCK}
\end{figure*}
\begin{figure}
    \centering
    \includegraphics[width=0.48\textwidth]{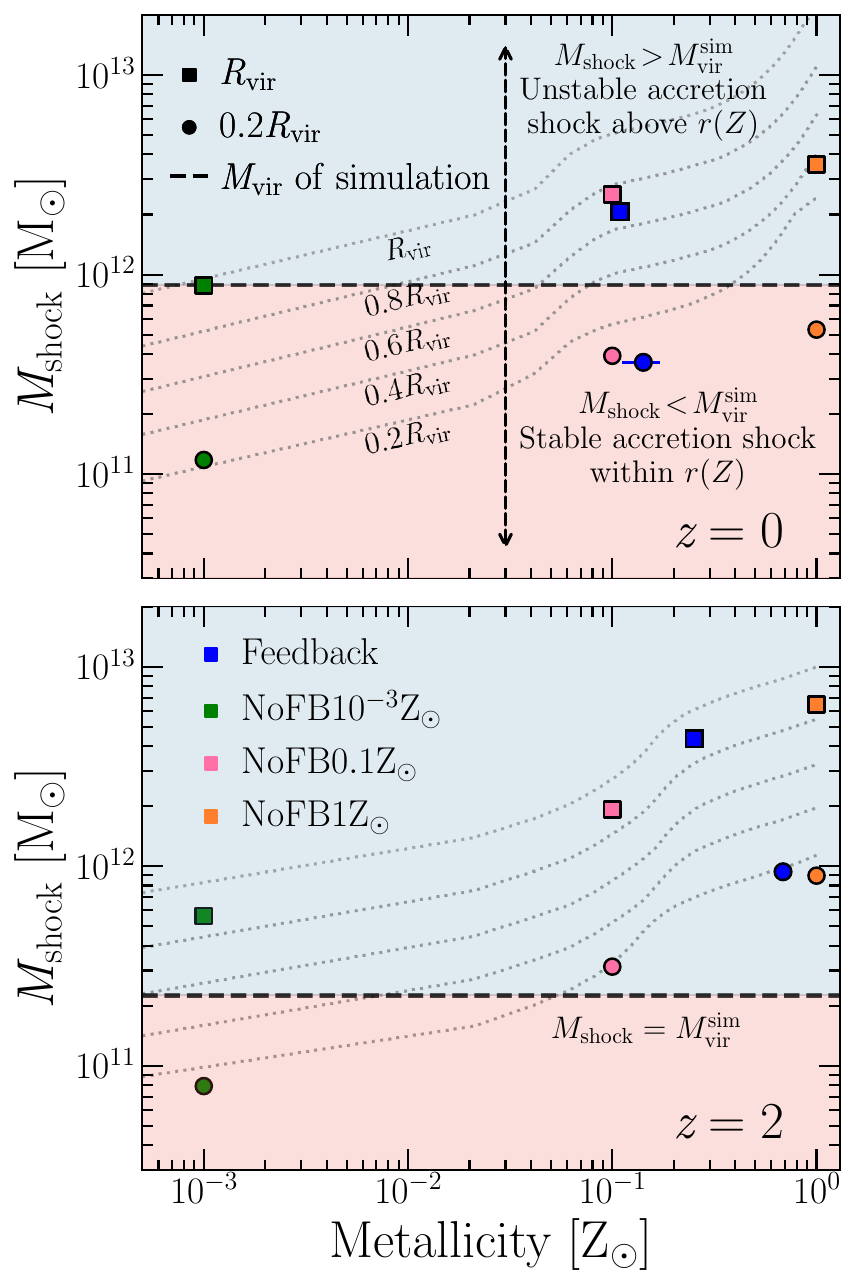}
    \caption{The expected $M_\text{shock}$ at a specific (mass-weighted median in our runs) metallicity. We compute the $M_\text{shock}$ at $R_{\text{vir}}$ (square symbol) and $0.2R_{\text{vir}}$ (circle symbol) for the Feedback, NoFB10$^{-3}Z_\odot$, NoFB$0.1Z_\odot$, and NoFB$1Z_\odot$ simulations at $z\!=\!2$ (bottom panel) and $z\!\sim\!0$ (upper panel). The horizontal lines are the $M_\text{vir}$ at each redshift. For a particular time, if $M_\text{vir}$ is above both the square and the circle symbols, the hot corona is established in the whole CGM, out to $R_{\rm vir}$. Only metal-poor haloes are able to establish the hot corona in the entire CGM, out to $R_{\rm vir}$, at $z\!=\!0$. To compare our results with the predictions of \protect\citetalias{Dekel2006}, we include, as thin, dotted lines, our calculations for different shells at fractions of $R_{\text{vir}}$ in the same range of metallicities and at redshift 0 as the original work of \protect\citetalias{Dekel2006}.}
    \label{FIG:MSHOCK_METALS}
\end{figure}

\begin{figure*}
    \centering
    \includegraphics[width=1\textwidth]{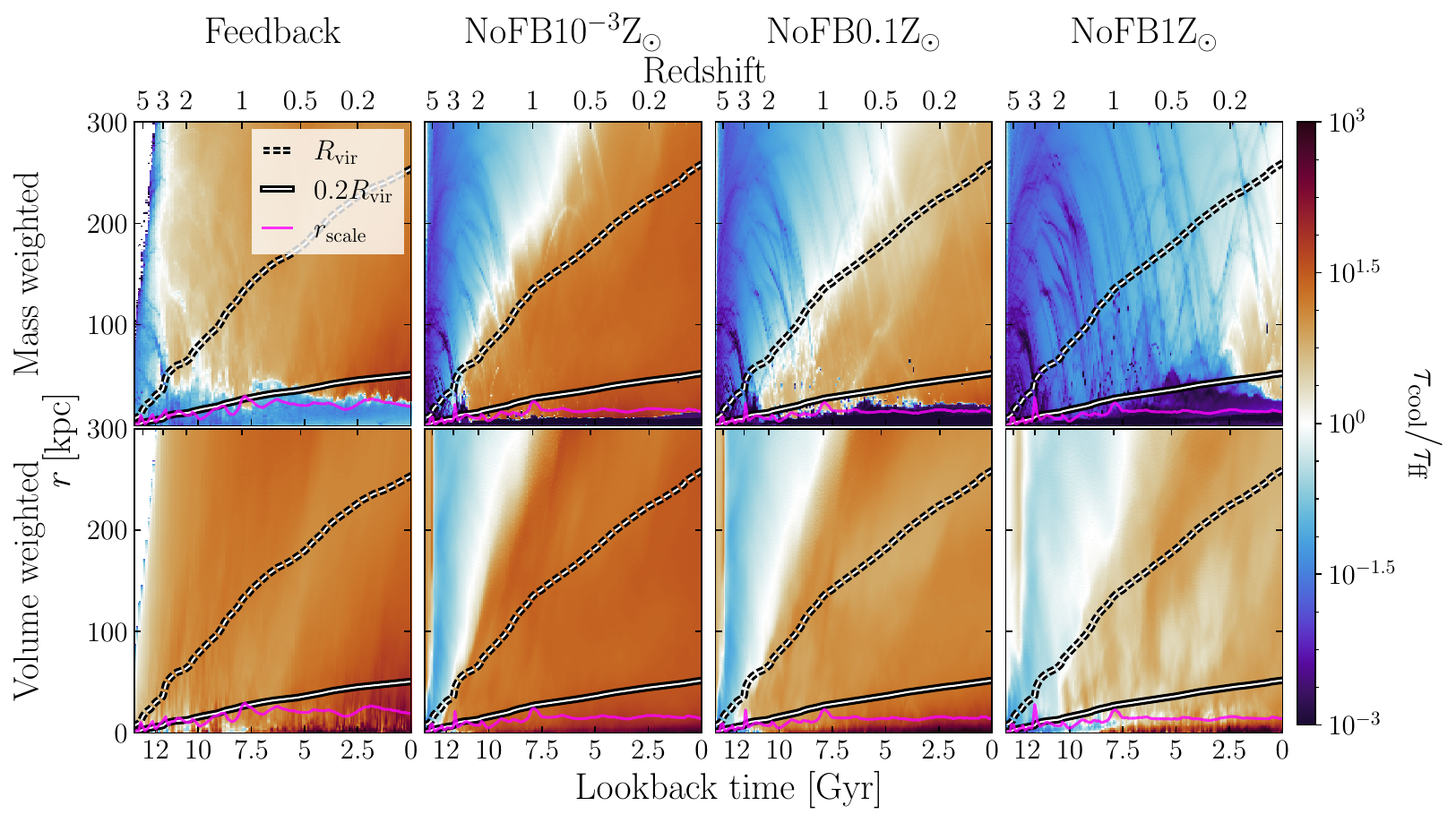}
    \caption{Top panels: mass-weighted $\tau_{\rm cool}/\tau_{\rm ff}$ ratio in radial bins within 300 kpc (proper distance) of the halo's centre ($y$-axis) as a function of time ($x$-axis) for our simulations (see also the first column in Fig. \ref{FIG:Vir_EarlyLate}). The layout and the meaning of the overlaid white curves are the same as in Fig.~\ref{FIG:HISTOGRAMS}. Regions with $\tau_{\rm cool}/\tau_{\rm ff}>1$ indicate virialised gas. Bottom panels: same as top panels but volume weighted.}
   
    \label{FIG:VIRIALIZATION_HIST}
\end{figure*}
\begin{figure*}
    \centering
    \includegraphics[width=0.9\textwidth]{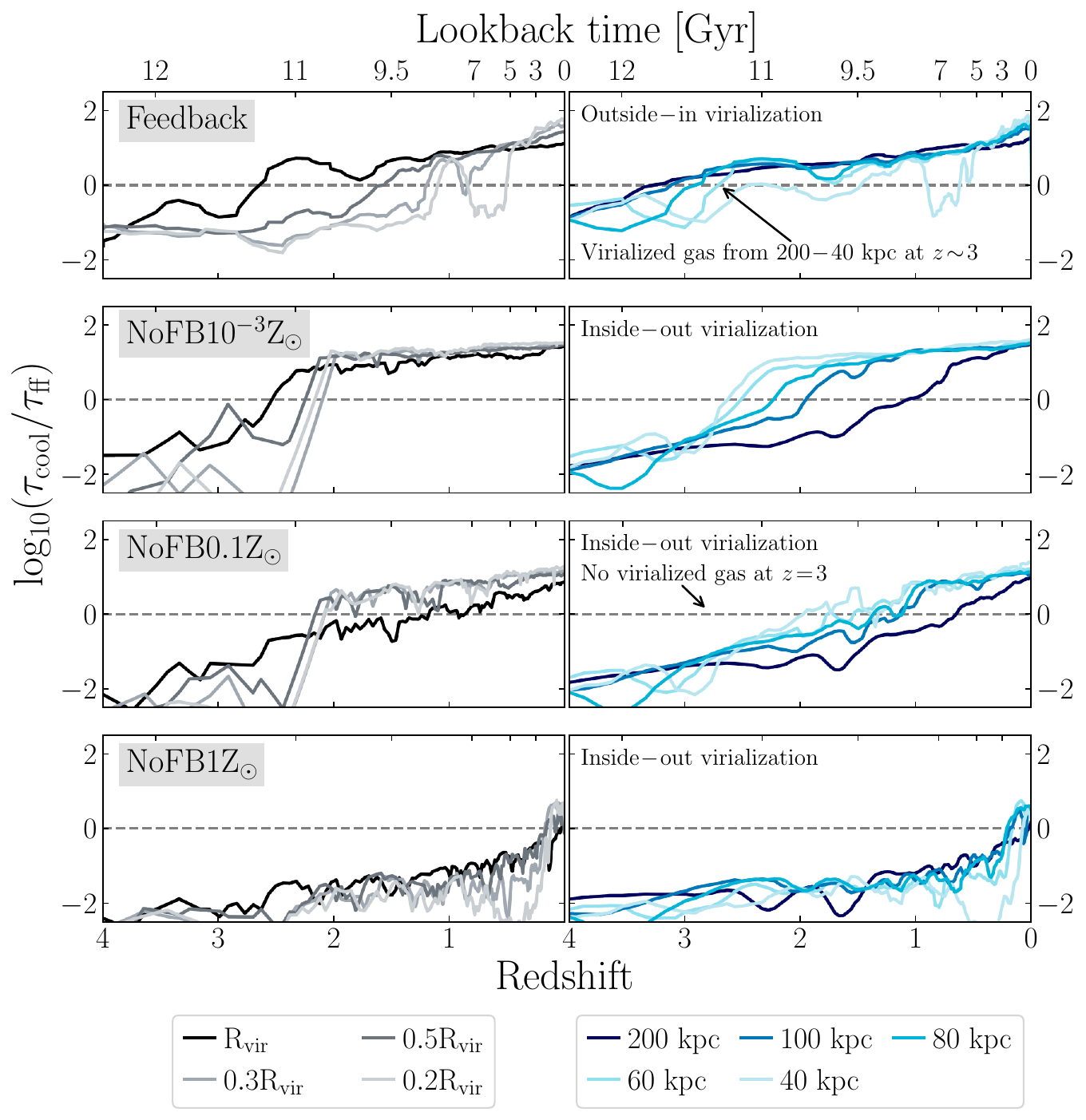}

    \caption{Evolution of the $\tau_{\text{cool}}/\tau_{\text{ff}}$ ratio within $R_{\text{vir}}$-based shells (left panel) and fixed physical-distance shells (proper scale, right panel).
}
    \label{FIG:SCALES}
\end{figure*}

\begin{figure}
    \centering
    \includegraphics[width=0.98\linewidth]{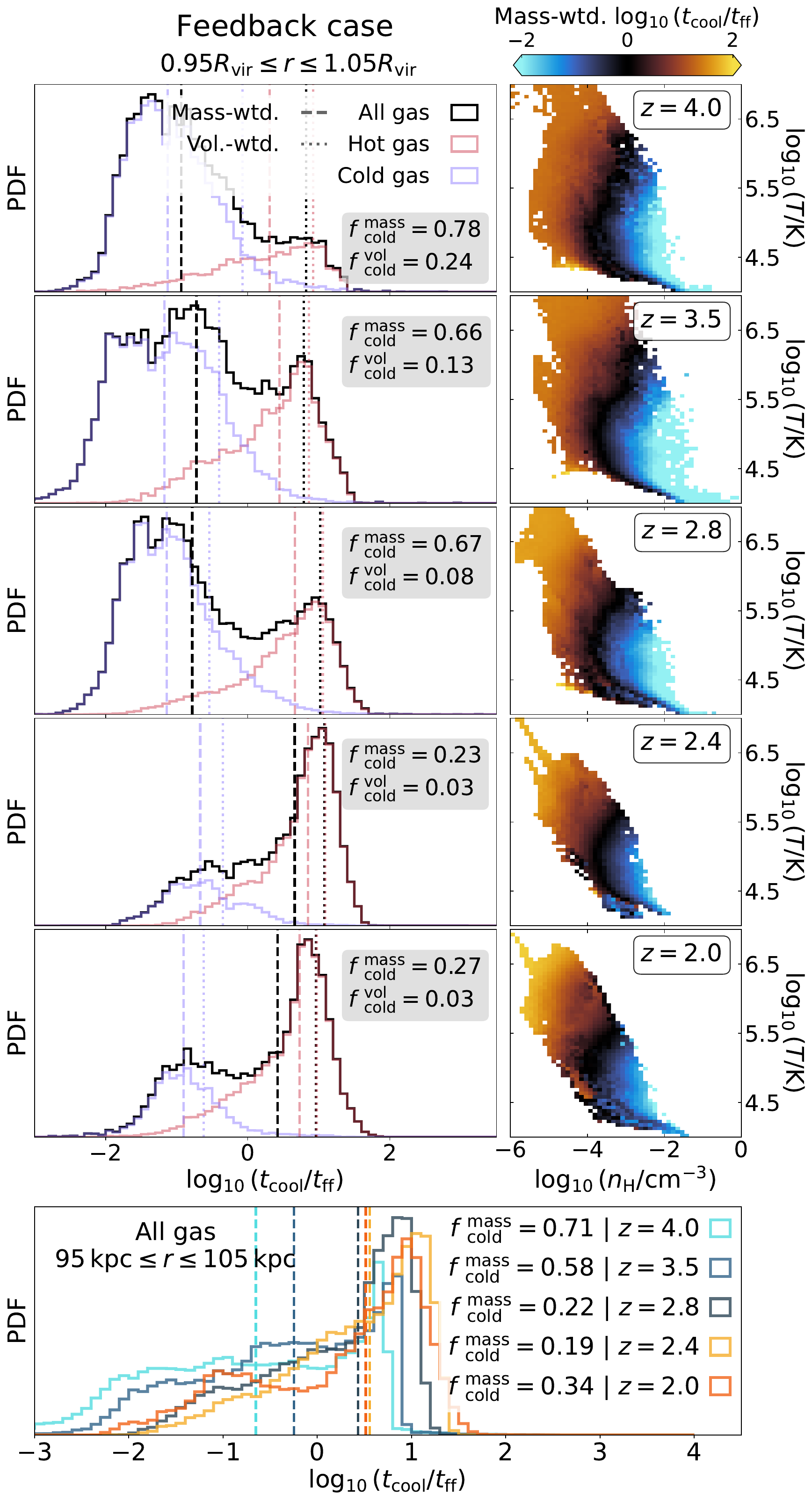}
\caption{Distribution of mass-weighted $\tau_{\rm cool}/\tau_{\rm ff}$ at different epochs around the transition to $\tau_{\rm cool}/\tau_{\rm ff}>1$, measured in the $R_{\rm vir}$ shell for the Feedback case (see Fig.~\ref{FIG:SCALES}). Left panels show the distributions for all, hot, and cold gas, with mass- and volume-weighted medians marked by dashed and dotted lines. Cold-gas mass and volume fractions are indicated in each panel. Right panels show the corresponding $T$--$n_{\rm H}$ phase diagrams, colored by $\log_{10}(\tau_{\rm cool}/\tau_{\rm ff})$. The bottom panel shows the evolution of the mass-weighted all-gas distribution at a fixed physical radius of $95$--$105\,{\rm kpc}$, with dashed lines marking the corresponding mass-weighted medians.}
    \label{fig:distirbution_tcooltff_rvir}
\end{figure}


\section{Results}
\label{sec:results}
In this section, we present the analysis of our Feedback and NoFB runs. Our goal is to address three questions: 1) When is the hot corona established? 2) What are the main heating mechanisms of the CGM gas for the hot corona formation? And 3) how does the onset of the hot corona relate to the virialisation process? We focus on the role of stellar feedback and metallicity, while the impact of different halo assembly histories is discussed separately in Appendix~\ref{sec:ap_AMH}.

\subsection{The formation of the hot corona in {\small VINTERGATAN}} \label{sec:inside_outHotCorona}
In this section, we explore the effect of stellar feedback on the formation and evolution of the hot corona, and how the feedback-driven process differs from the pure accretion shock scenario.

In Fig.~\ref{FIG:MULTITEMP}, we show gas temperature maps for the Feedback (top row) and NoFB$0.1Z_\odot$ (bottom row) runs at four selected cosmological epochs ($z = 4, 3, 2$ and 0). In the Feedback run, gas within, and beyond, $R_{\rm vir}$ (indicated by the dashed circle in each panel) undergoes rapid heating and reaches temperatures above $10^5\,\mathrm{K}$ as early as $z \gtrsim 4$ (top-left). In contrast, when stellar feedback is absent and accretion shocks are the sole heating source, the hot gas component is less extended, largely confined to well within $R_{\rm vir}$ at $z \gtrsim 2$, and only at later times (lower $z$) does it expand beyond $R_{\rm vir}$. This difference shows that stellar feedback is the primary mechanism driving the early onset and rapid expansion beyond $R_{\rm vir}$ of the hot corona in our runs.

This comparison is particularly relevant because the NoFB$0.1Z_\odot$ run has a metallicity that is representative of the Feedback run (see Fig.~\ref{FIG:Metallicity_evolving}). As a result, the two runs have similar metal-dependent cooling efficiencies at fixed density and temperature, although their actual CGM cooling rates may still differ because of differences in the gas density. This allows a cleaner isolation of the respective roles of accretion shocks and stellar feedback in driving the onset of the hot corona. For a more detailed comparison of the impact of varying halo metallicity, see Fig.~\ref{FIG:MultuTemp_NoFBrest} and Sec. \ref{sec:metal_dependency}.

In order to show the full evolution of the hot corona within its host halo, left (right) panels in Fig. \ref{FIG:HISTOGRAMS} shows the mass (volume)-weighted column temperature in spherical radial bins within 300 kpc (proper distance) from the halo's centre, as a function of time and for the same runs as shown in Fig. \ref{FIG:MULTITEMP}. We include $R_{\rm vir}$ and $0.2R_{\rm vir}$ as white dashed and solid lines, respectively, to indicate our adopted CGM boundaries. The evolution of the CGM geometry is discussed further in Sec.~\ref{sec:OutInVir_Pseudo}.

The top panels illustrate how feedback rapidly injects energy into the gas over large distances and ejects hot material, visible as red ``feathers" extending from $\sim$50 kpc to 300 kpc in all panels. At $z\!<\!3$, most gas, except in the innermost regions, is heated to temperatures close to the virial temperature, producing a very sudden transition from a cold to a hot CGM at \emph{all} distances (out to many virial radii) beyond $\sim 30-50$ kpc. This is visible in all panels, but more dramatically in the volume-weighted one, where the volume-filling gas presents $T>10^5\ {\rm K}$ at even earlier times. As such, heating by feedback does not produce a clear inside-out formation scenario for the hot corona. In contrast, the bottom panels showing the NoFB run reveal that accretion shocks can be seen to reach the outer regions more slowly, producing a clearer inside-out pattern and a ``cone-shaped" transition between cold, bluish region to the hot, reddish region, consistent with an expanding hot corona via accretion shock. This is visible for both panels, mass-weighted and volume-weighted.

\subsection{Metallicity dependence of hot corona's onset}
\label{sec:metal_dependency} 
In the previous section, we showed that stellar feedback promotes the early formation of a hot corona and drives hot gas out to several $R_{\rm vir}$, whereas accretion shocks heat gas more slowly and mainly within $R_{\rm vir}$. However, as discussed in Sec.~\ref{sec:virshock}, the efficiency of accretion-shock heating depends on gas metallicity. This dependence raises the question of whether variations in the halo metal content could allow accretion shocks to reproduce the rapid, early growth of the hot corona seen in the Feedback run. In this section, we test whether stellar feedback remains the primary mechanism driving the early, halo-wide formation of the hot corona when we vary the metallicity in the NoFB runs, namely NoFB$10^{-3}Z_\odot$, NoFB$0.1Z_\odot$, and NoFB$1Z_\odot$.

To quantify the different evolutionary paths of the hot corona, the top and bottom panels of Fig.~\ref{FIG:TVIR_TatRVIR} show the gas mass-weighted median temperature in a shell centred at $R_\text{vir}$ as a function of $M_\text{vir}$ (right panels), and compare it with the theoretical $T_\text{vir}$ (left panels), defined in Eq.~\ref{eq:tvir}. The top panels show the mass-weighted temperature, which traces the gas phase that dominates the mass in this shell, while the bottom panels show the volume-weighted temperature, which traces the volume-filling gas.

In the mass-weighted panels, at $z\!\lesssim\!4$, stellar feedback raises the gas temperature at $R_\text{vir}$ to values close to $T_\text{vir}$. The NoFB runs do not show this behavior for any of the adopted metallicities; instead, their temperatures remain below $T_\text{vir}$ and differ from the Feedback run by $\sim 0.2$--$1$ dex. This offset shows that varying the metal content does not allow accretion shocks alone to reproduce the early heating seen in the Feedback run. The volume-weighted temperature shows an even stronger contrast between the Feedback and NoFB runs. When feedback is included, the volume-filling gas reaches values close to $T_\text{vir}$ before $z\sim4$. In the NoFB runs, however, the volume-weighted trends are similar to the mass-weighted ones.

Among the NoFB runs, higher metallicity enhances radiative cooling, making the discrepancy with the Feedback run larger as the metal content increases. However, NoFB$10^{-3}Z_\odot$ and NoFB$0.1Z_\odot$ follow similar trends at $z\!<\!2$ for both mass- and volume-weighted temperatures. In addition, the right panels show that once the halo reaches $M_{\rm vir}\sim10^{11.2}\,{\rm M_\odot}$, the temperature at $R_{\rm vir}$ remains close to $\sim10^{5.5}$ K in all cases (except NoFB$1Z_\odot$) with no significant subsequent evolution. This roughly constant temperature suggests that the virial temperature at this radius evolves only weakly after this point, consistent with a gravitational potential that has largely settled. We further discuss this last result in Sec.~\ref{sec:OutInVir_Pseudo}.

\subsection{Expected hot corona formation in {\small VINTERGATAN} under the Virial Shock model scenario}\label{sec:expectedOnset} 
Having shown that accretion shocks alone, even considering large metallicity variations, cannot account for the early formation of the hot corona in the Feedback simulation, we now use the Virial Shock model  \citepalias[][see Sec.~\ref{sec:virshock}]{Birnboim2003MNRAS.345..349B, Dekel2006} to assess when a hot corona would emerge if virial shock heating were the only mechanism responsible for maintaining the thermal state of the CGM. To do so, we take the halo mass growth and CGM metallicity evolution directly from the Feedback simulation and use them as inputs to Eq.~\ref{eq:Mshock}. This procedure isolates the prediction of the virial-shock framework while retaining the assembly history and metal enrichment produced by stellar feedback. Fig.~\ref{FIG:MSHOCK} shows the resulting evolution of the critical shock mass, $M_{\rm shock}$, required to sustain a stable accretion shock at $R_{\rm vir}$ and $0.2R_{\rm vir}$. The evolution of $M_{\rm vir}$ is also shown as a black line, indicating when the stability condition $M_{\rm vir} > M_{\rm shock}$ is satisfied.

Using the halo growth and metallicity history of the Feedback simulation, the Virial Shock model predicts an inside-out formation of the hot corona. At all epochs, $M_{\rm shock}$ at $0.2R_{\rm vir}$ remains lower than at $R_{\rm vir}$, implying that a stable shock first develops in the inner CGM before propagating outward. According to the model, the inner CGM becomes susceptible to sustained shock heating only at $z \lesssim 1.5$, while a stable virial shock at $R_{\rm vir}$ is not established by $z=0$. These predictions differ markedly from the thermal evolution measured in the simulation (Figs.~\ref{FIG:HISTOGRAMS} and \ref{FIG:TVIR_TatRVIR}), where a hot corona is already in place at substantially earlier times. Since the Virial Shock model is supplied with the same halo growth and metal enrichment history as the simulation, this discrepancy demonstrates that these factors alone are insufficient to explain the emergence of the hot CGM. Instead, direct heating by stellar feedback must play a dominant role in establishing the thermal state of the circumgalactic medium.

In the NoFB runs, a stable shock is predicted to form in the inner CGM at markedly different epochs, compared to the FB run, with lower metallicities leading to earlier transitions. This contrasts with the similar temperature evolution at $R_{\rm vir}$ shown in Fig.~\ref{FIG:TVIR_TatRVIR}, where NoFB$10^{-3}Z_\odot$ and NoFB$0.1Z_\odot$ exhibit comparable temperature trends at $z<2$.

To better understand the origin of this metallicity dependence, Fig.~\ref{FIG:MSHOCK_METALS} shows how the stability of accretion shocks varies with metallicity at $z=0$ and $z=2$. We evaluate the model at two radii, $0.2R_{\rm vir}$ (circle symbols) and $R_{\rm vir}$ (square symbols). The figure therefore illustrates the maximum radius at which a stable accretion shock can be sustained as a function of metallicity, i.e. $r=r(Z)$, for a halo of mass $M_{\rm vir}^{\rm sim}$.

At $z=0$ (top panel), and for metallicities around $0.1Z_\odot$, comparable to those reached in the Feedback run, the accretion shock is not stable at $R_{\rm vir}$ but is still expected to support the growth of the hot corona beyond $0.2R_{\rm vir}$, under the assumptions of the accretion-shock theory proposed in \citetalias{Birnboim2003MNRAS.345..349B,Dekel2006}. The Feedback and NoFB$0.1Z_\odot$ runs show similar behaviour, suggesting that, at fixed halo formation history and metallicity, the model predicts comparable shock stability with and without stellar feedback, as feedback is not included by construction in \citetalias{Birnboim2003MNRAS.345..349B, Dekel2006}'s framework.

At $z=2$ (bottom panel), the shock is unstable at $R_{\rm vir}$ in all but the most metal-poor case, NoFB$10^{-3}Z_\odot$. The model therefore predicts that a stable hot corona should not exist at such early epochs except under extremely low-metallicity conditions. Unlike the situation at $z=0$, the Feedback and NoFB$0.1Z_\odot$ runs no longer show similar behaviour at $z=2$: the Feedback run is less likely to develop a stable virial shock because metal enrichment increases the cooling efficiency. This prediction is inconsistent with the real simulated evolution, where the Feedback run develops the hot corona significantly earlier than any of the NoFB models (Fig.~\ref{FIG:TVIR_TatRVIR}).

While the Virial Shock model correctly captures the qualitative trend that lower metallicities favour earlier hot-corona formation, the simulations show that hot gas extends to approximately $R_{\rm vir}$ by $z\sim2$ in most NoFB runs. This suggests that this 1D model overpredicts the threshold mass required for stable shock heating in Milky Way-mass halo progenitors. 

\subsection{The virialisation of the CGM in {\small VINTERGATAN}}\label{sec:outside-inVIR}

In this section, we examine how CGM virialisation proceeds in the {\small VINTERGATAN} suite, using $\tau_\text{cool}/\tau_\text{ff} > 1$ as our virialisation criterion. Our goal is to determine whether virialisation follows the same evolutionary trends as the development of a hot corona (Sec. \ref{sec:inside_outHotCorona}), and how it is affected by both stellar feedback and metallicity.

Top (bottom) panels in Fig.~\ref{FIG:VIRIALIZATION_HIST} shows the mass (volume)-weighted $\tau_{\rm cool}/\tau_{\rm ff}$ ratio in radial bins as a function of time within 300 kpc (proper distance) of the halo's centre for our simulations. In the Feedback and NoFB$0.1Z_\odot$ runs, regions with $\tau_{\rm cool}/\tau_{\rm ff}>1$ match the hot coronal gas regions identified in Fig.~\ref{FIG:HISTOGRAMS}. This strongly suggests that CGM virialisation begins once the hot corona is established. As shown in Sec.~\ref{sec:inside_outHotCorona}, stellar feedback accelerates the $\tau_{\rm cool}/\tau_{\rm ff}>1$ transition compared to the NoFB runs, producing a rapid change from non-virialised to virialised gas across almost the full 300 kpc region considered here. In the NoFB runs, the virialisation occurs systematically later as the metallicity increases. NoFB$1Z_\odot$ stands out as an extreme case, where the gas remains non-virialised at all radii until $z\lesssim0.2$ when $\tau_{\rm cool}/\tau_{\rm ff}$ is mass-weighted, reflecting the large contribution of cold and dense gas. When $\tau_{\rm cool}/\tau_{\rm ff}$ is volume-weighted, the ratio reveals extended regions spanning several hundred kpc that fluctuate around unity, indicating a CGM close to the transition between efficient and inefficient cooling. It therefore remains in an unstable transitional state rather than establishing a persistently virialised, volume-filling phase.

The next question is how this virialisation front propagates through the halo. To quantify this, Fig.~\ref{FIG:SCALES} shows the evolution of the mass-weighted median of $\tau_{\text{cool}}/\tau_{\text{ff}}$ in individual radial shells. The left column corresponds to shells defined in units of $R_\text{vir}$, following the standard geometrical definition of the CGM \citep[e.g.][]{Stern2021,Correa2018MNRAS.473..538C,Hopkins2023}; we address the limitations of this definition in later sections. We also remark here that this diagnostic traces the thermodynamic state of the multiphase CGM rather than directly identifying when a long-lived hot corona is established, and can therefore also reflect changes in the dominant gas accretion mode.

In the Feedback run, the condition that $\tau_{\text{cool}}/\tau_{\text{ff}}\!>\!1$ is reached in an outside-in manner, consistent with previous work \citep{Stern2021,Sultan2024arXiv241016359S}\footnote{We
emphasize that \protect\citealt{Stern2021,Sultan2024arXiv241016359S} primarily characterize the virialisation
of the volume-filling hot phase, whereas our mass-weighted analysis
includes the full multiphase CGM. See Sec. \ref{sec:vir_distirbution}.}. The transition from $R_\text{vir}$ down to $0.2R_\text{vir}$ takes $\sim5$ Gyr. By contrast, the NoFB runs show a different behaviour. For NoFB$0.1Z_\odot$ and NoFB$1Z_\odot$, virialisation first occurs within $\sim0.5R_{\rm vir}$, while the shell at $R_{\rm vir}$ is the last to virialise, although in the latter case this happens shortly after the inner shells reach $\tau_{\text{cool}}/\tau_{\text{ff}}\!>\!1$. This can be broadly described as inside-out virialisation. NoFB$10^{-3}Z_\odot$ follows a similar evolution to NoFB$0.1Z_\odot$, but reaches $\tau_{\text{cool}}/\tau_{\text{ff}}\!>\!1$ at $R_{\rm vir}$ earlier, and before the inner shells. This first analysis using $R_{\rm vir}$-based shells suggests that virialisation within $\sim0.5R_{\rm vir}$ occurs faster than in the Feedback run, although not necessarily at earlier cosmological epochs, and does not necessarily follow an outside-in pattern. This difference can be partly explained by the effect of feedback on the gas structure: feedback produces a stratified metallicity distribution and efficiently heats gas and ejects hot gas at large radii, while in the NoFB runs the metallicity is fixed and no feedback heating is present. As a result, stellar feedback can virialise the outer halo while a colder, metal-rich phase persists in the inner CGM.

However, as discussed in the previous section and shown in Fig.~\ref{FIG:HISTOGRAMS}, once the hot corona is established at $z\!\sim\!2-3$, the temperature and $\tau_{\rm cool}/\tau_{\rm ff}$ of the gas within 300 kpc evolve only weakly, while $R_{\rm vir}$ continues to grow. Therefore, trends inferred from $R_{\rm vir}$-scaled shells may partly reflect the evolving CGM boundary rather than the underlying thermal or dynamical state of the CGM gas. To test this, the right column of Fig.~\ref{FIG:SCALES} repeats the analysis using fixed physical shells. In this approach, the Feedback run still shows an outside-in transition, but over a much shorter time interval, $\sim 0.5$ Gyr, recovering the rapid, halo-wide transition to a hot and virialised gas due to stellar feedback. As commented, the fixed physical shells also clarify the inside-out behaviour of the NoFB runs and emphasises the strong metallicity dependence of their virialisation times: the most metal-rich case virialises last because enhanced cooling delays the transition to $\tau_{\rm cool}/\tau_{\rm ff}\!>\!1$. This suggests that the growth of $R_{\rm vir}$ contributes to the apparent timing of virialisation in $R_{\rm vir}$-scaled shells. We discuss this point further in Sec.~\ref{sec:OutInVir_Pseudo}.
  
Finally, the fixed-kpc analysis makes the impact of stellar feedback on halo virialisation particularly evident. Comparing the Feedback and NoFB$0.1Z_\odot$ runs, the Feedback run already contains virialised gas over $\sim40$--$100$ kpc by $z\!=\!3$. In the corresponding NoFB run, a comparable state is not reached until $z\!=\!1$. Thus, stellar feedback can both change the direction of virialisation and accelerate the transition to a virialised CGM by several Gyr. This allows virialisation to occur nearly simultaneously across a large volume of the halo. This transition has direct implications for the expected mode of gas accretion. Once the CGM satisfies $\tau_{\rm cool}/\tau_{\rm ff}>1$, gas accreted from the CGM onto the galaxy is expected to originate from a pressure-supported hot atmosphere and flow inward through radiative cooling. We therefore next ask whether the early virialisation measured in our simulations is consistent with the stability threshold predicted by the Cooling Flow model. In Appendix \ref{sec:ap_coolingflow}, we perform this comparison using halo and CGM properties measured directly from the simulations as inputs to the model.

\subsection{The role of hot and cold gas phases in CGM virialisation}\label{sec:vir_distirbution}
In this section, we examine in detail how CGM virialisation proceeds through the evolution of the cold ($T_{\rm gas}<10^5\,{\rm K}$) and hot ($T_{\rm gas}>10^5\,{\rm K}$) gas phases, and how this evolution is reflected in mass- and volume-weighted statistics. We focus on the Feedback case as a reference, since it undergoes the earliest transition to a virialised CGM, and use the $R_{\rm vir}$ shell as a representative probe of the CGM, see Figure \ref{FIG:VIRIALIZATION_HIST} for the evolution of $R_{\rm vir}$.

Figure~\ref{fig:distirbution_tcooltff_rvir} reveals a clear separation between the thermodynamic states of the hot and cold CGM phases; a bimodality in the $\tau_{\rm cool}/\tau_{\rm ff}$ distribution. At high redshift, the cold dense gas predominantly occupies the low-$\tau_{\rm cool}/\tau_{\rm ff}$ branch, while the hot phase already lies mainly at $\tau_{\rm cool}/\tau_{\rm ff}>1$ and shows little evolution thereafter. The evolution of the all-gas mass-weighted median is therefore largely driven by the changing contribution of the cold component: at $4>z\gtrsim2.8$, cold dense gas accounts for $\sim65$--$80\%$ of the mass in the shell, causing the mass-weighted median to follow the low-$\tau_{\rm cool}/\tau_{\rm ff}$ branch. By $z\simeq2.4$, its mass fraction drops to $\sim\!23\%$, and the all-gas median rapidly approaches that of the hot phase. Interestingly, this decrease approximately coincides with the last major merger experienced by the halo; whether the two events are physically connected will be explored in future work. In contrast, the cold gas occupies only a small fraction of the  volume throughout, falling from $\sim24\%$ at $z=4$ to only $\sim3\%$ at $z\lesssim2.4$, such that the volume-weighted statistic provides closely matches that of the volume-filling hot component. This demonstrates that, as shown in Figs. \ref{FIG:HISTOGRAMS} and \ref{FIG:VIRIALIZATION_HIST} for the Feedback case, a hot corona can already be established at high redshift even when it represents only a minor fraction of the CGM mass, falling to $\lesssim20\%$ at $z>4$, consistent with stellar feedback promoting its early formation. 

Following the same argument as in Sec.~\ref{sec:outside-inVIR}, based on the finding that the thermodynamic properties of the CGM beyond a few tens of kpc evolve only weakly once the hot corona is established, we repeat the $\tau_{\rm cool}/\tau_{\rm ff}$ distribution analysis in a fixed $95$--$105\,{\rm kpc}$ shell. Bottom panel in Fig. \ref{fig:distirbution_tcooltff_rvir} shows that the cold-gas contribution declines earlier, and the $\tau_{\rm cool}/\tau_{\rm ff}$ distribution subsequently evolves only weakly at $z\lesssim3.5$. This is consistent with the earlier behaviour seen in Figs.~\ref{FIG:HISTOGRAMS}, \ref{FIG:VIRIALIZATION_HIST}, and \ref{FIG:SCALES}, where the hot-CGM properties at these radii are established earlier. Hence, the transition to $\tau_{\rm cool}/\tau_{\rm ff}>1$ is driven by the decline of the dense cold-gas fraction.

The $T$--$n_{\rm H}$ diagrams in the right panels reinforce this picture: the densest gas is predominantly cold and is associated with lower $\tau_{\rm cool}/\tau_{\rm ff}$, whereas hotter and more diffuse gas generally occupies larger ratios. Nevertheless, in all panels it is possible to find a small amount of cold (hot) gas extending to relatively large (short) $\tau_{\rm cool}/\tau_{\rm ff}$, although it remains marginal in the overall distribution. This overlap highlights the genuinely multiphase nature of the CGM and shows that neither a simple temperature cut nor a single mass- or volume-weighted statistic can fully characterize its thermodynamic structure.

A related point was discussed by \citet{Stern2021} and \citet{Kakoly2025arXiv250417001K}, who showed that the instantaneous cooling time measured directly from simulated gas does not always provide a unique indicator of whether that gas belongs to a long-lived virialized phase. In particular, gas that has already cooled can subsequently exhibit a relatively long $t_{\rm cool}$, providing a possible interpretation for the small population of cold gas with large $\tau_{\rm cool}/\tau_{\rm ff}$ seen here. More generally, \citet{Esmerian2021} showed that $t_{\rm cool}/t_{\rm ff}$ itself can be a poor predictor of thermal stability in gaseous haloes when large-amplitude density perturbations are present, since gas parcels with similar $t_{\rm cool}/t_{\rm ff}$ can have very different subsequent cooling histories. 


\section{Discussion}\label{sec:DISCUSSION}

\subsection{Why is the hot corona established so early in {\small VINTERGATAN}? } \label{sec:Dis_earlycorona}
In Sec.~\ref{sec:expectedOnset} and Sec.~\ref{sec:inside_outHotCorona}, we showed that the Virial Shock model, represented here by the pure accretion--shock scenario of the NoFB runs, cannot reproduce the early onset of the hot corona at $z\!\sim\!3$ found in the simulations with stellar feedback. This result does not place our simulations in conflict with previous numerical studies. Instead, it highlights a difference in interpretation. While the overall evolution of the CGM is broadly consistent with the literature, most previous works do not focus on identifying when the hot corona first appears, nor on separating the relative roles of virial accretion shocks and feedback-driven heating.
  
Our results suggest that, in Milky Way--mass haloes, the early hot corona is established primarily by stellar feedback. Feedback-driven shocks propagate rapidly beyond galactic scales, reaching distances of up to $\sim\!300$ kpc in our simulations. This hot phase is maintained until $z\!=\!0$. As highlighted by \cite{Faucher2023ARA&A..61..131F}, while such large-scale heating by feedback can be seen in simulations such as {\small EAGLE} \citep{Correa2018MNRAS.473..538C, Wijers2020MNRAS.498..574W}, {\small FIRE-2} \citep{Ji2021MNRAS.505..259J}, with comparative simulation studies such as \cite{Wright2024}, and the AGORA analysis of \cite{StrawnAGORA2024ApJ...962...29S}, it is not always emphasized or demonstrated to be an outcome of feedback heating. A recent exception is  \citealt{messere2026} who demonstrated that for dwarf galaxies forming in $10^{10}-10^{11}~{\rm M_\odot}$ haloes, equivalent to our simulations at $z>3$, SNe-driven shock heating can sustain a hot and highly virialized CGM.

The presence of this early, feedback-heated corona does not imply that the inner CGM becomes fully hot. Instead, it is consistent with a long-lived multiphase medium, where a volume-filling hot atmosphere coexists with colder and denser gas near the galaxy, as suggested by observations \citep[e.g.][]{Schrijver2001SoPh,Fabian2003MNRAS.344L..48F, Putman2012ARA&A..50..491P} and by numerical studies \citep[e.g.][]{Sharma_2010,Waters_2019,Schneider_2020,Das2021MNRAS.502.4935D,Dutta2022MNRAS.510.3561D,Kere2005,Kere2009MNRAS.395..160K,Kere2009,Ocvirk2008MNRAS.390.1326O,Khalatyan2008MNRAS.387...13K,Voort2011}.

This distinction is important when comparing with studies that associate hot-corona formation with the transition from cold-to-hot mode dominated accretion. In those works, the hot corona is often linked to the stage at which hot accretion becomes dominant, typically around $M_{\rm vir}\!\sim\! M_{\rm thres}$, with $M_{\rm thres}\sim 10^{11-12}\ {\rm M}_{\odot}$ in qualitative agreement with \citetalias{Dekel2006}. However, it is not always clear from the literature whether the hot phase develops inside-out or outside-in, or which heating source dominates its formation.

In our simulations with stellar feedback, the hot corona appears before the halo reaches the well-known $M_{\rm thres}$ regime of $10^{11.5}$--$10^{12}\,M_\odot$ \citep{Kere2005,Kere2012,Nelson2013}. In Fig. \ref{FIG:TVIR_TatRVIR}, we showed that by $z\!\sim\!3$, when $M_{\rm vir}\!\lesssim\!10^{11.5}\,M_\odot$, the mass-weighted temperature at $R_{\rm vir}$ already satisfies $T(R_{\rm vir})\sim T_{\rm vir}$. This transition occurs even earlier in the volume-weighted temperature: the volume-filling gas reaches $T_{\rm vir}$ already at $z\!>\!3.5$, as shown in Fig.~\ref{FIG:TVIR_TatRVIR}. The final stabilization of $T(R_\text{vir})$ is reached at $M_\text{vir}\!\sim\!10^{11.2}\,M_\odot$. We note, however, that $T(R_{\rm vir}) \sim T_{\rm vir}$ should not be taken as a necessary condition for virialisation: in the {\small FOGGIE} simulations \citep{Peeples2019}, halo gas can remain cooler than $T_{\rm vir}$ while staying close to virial equilibrium, because non-thermal turbulent and bulk motions provide part of the dynamical support \citep{Lochhaas2021, Lochhaas2023ApJ...948...43L}. In our simulations, the feedback-driven heating at large radii also appears to be independent of halo metallicity and halo-mass assembly history, as suggested by Figs.~\ref{FIG:Vir_EarlyLate} and \ref{FIG:Metallicity_evolving}. This is likely due to weak metal mixing between shells, which produces similar gas profiles in the outer CGM. We will explore this point further in upcoming work.
  
The NoFB runs provide a useful contrast because they isolate the accretion--shock's role. In these simulations, a hot corona can still form without feedback, but only on longer timescales and through a more gradual inside-out process. Overall, the NoFB runs follow the qualitative behavior expected from the Virial Shock model: higher metallicity increases the cooling efficiency and delays the establishment of the hot corona. This trend is consistent with the sensitivity of the model to the cooling function \citepalias{Dekel2006}, and with previous work showing that chemical enrichment plays an important role in the transition from cold to hot diffuse gas accretion \citep{Ocvirk2008MNRAS.390.1326O}.
  
However, the comparison is not fully quantitative. The analytical model tends to overpredict the halo mass required to establish a hot corona at $R_{\rm vir}$ through accretion shocks. In other words, the NoFB simulations develop hot gas at large radii earlier, or at lower halo masses, than predicted by the idealized model. This difference is likely related to the assumptions entering the analytical calculation. For example, in the Virial Shock model, the pre-shock baryon density is estimated from an averaged cosmological density prescription, $\rho_b = 4 f_{\rm b} (\rho/\bar{\rho})_{\rm vir}\Delta \rho_{\rm u} f_{\rm\rho}$\footnote{In this expression, $\rho_b$ is the pre-shock baryon density, $f_b$ is the effective baryonic fraction, $(\rho/\bar{\rho})_{\rm vir}$ converts the mean density inside $R_{\rm vir}$ into the local density at $R_{\rm vir}$, $\Delta$ is the mean overdensity within the virial radius, $\rho_u$ is the universal mean mass density, and $f_\rho \equiv \rho(r)/\rho(R_{\rm vir})$ accounts for the radial variation of the gas density relative to its value at the virial radius. The factor 4 accounts for the density jump between the pre-shock and post-shock gas.} \citepalias[Eq.~20]{Dekel2006}. This provides a characteristic density at the virial radius, but it does not capture the full multiphase and non-spherical structure of the simulated CGM. The low-density gas, which is heated more easily, is already present in the inner CGM, as shown in previous simulations \citep{Esmerian2021,Wright2024}. At high redshift, the CGM can also show a broad temperature distribution \citep{Wright2024} or comparable mass fractions between cold and hot phases \citep{Huscher2021MNRAS.500.1476H}.
  
Taken together, these results suggest that stellar feedback can strongly modify, and in some cases hinder, the development of stable accretion shocks. In the inner CGM, feedback changes the conditions for shock stability in two main ways. First, it introduces pressure gradients and density contrasts that can trigger hydrodynamic and thermal instabilities. Second, it transports metals from the ISM into the CGM, increasing the cooling efficiency if mixing is efficient. The impact of feedback on shock stability was examined by \cite{Fielding2017}, who showed that haloes below the critical mass of $10^{11.5}M_\odot$ have CGM properties dominated by feedback and by its interaction with inflowing gas. In our simulations, this mass scale roughly corresponds to the final stabilization of the temperature at $R_\text{vir}$, shown in Fig.~\ref{FIG:TVIR_TatRVIR}. Before this stage, $T(R_\text{vir})$ differs strongly between the Feedback and NoFB runs, indicating that feedback dominates the thermodynamic state of the CGM. Similar disruptions of accretion shocks have also been reported in simulations including cosmic-ray feedback \citep{Ji2021MNRAS.505..259J}. This supports the interpretation that accretion shocks are not the dominant mechanism setting the early onset of the hot corona in the Feedback run.

\subsection{Outside-in virialisation or pseudo-evolution?} \label{sec:OutInVir_Pseudo}
Although the formation of a hot corona and the virialisation of the CGM, the dynamical state of the gas with respect to cooling efficiency, do not in principle need to be connected, we find that both occur at nearly the same time, with no significant temporal offset. As such, the two concepts are often used without distinction in the literature. Interestingly, analytical models such as the Cooling Flow model \citep{Stern2019, Stern2020} describe that the inner CGM is the last region to present a sustained long-lived hot phase when evaluating $\tau_{\rm cool}/ \tau_{ff}$ for shocked hot gas, leading to an outside-in scenario for the virialization by construction (see sec. 2.3 in \citealt{Stern2019}). This is in opposite direction of both cosmological accretion and feedback-driven shocks, then the formation of the hot corona would also be expected to follow this trend (see Sec. \ref{sec:ap_coolingflow} for more details in our cooling flow calculations). This is somewhat counterintuitive, as in the simple scenario described by \citetalias{Dekel2006}, the heating process follows the accretion shock, i.e. in an inside-out manner.

This interpretation also depends on the statistic and method used to characterize the multiphase CGM, as shown in the appendix~A of \citealt{Stern2021}. The cold dense component can dominate the mass while filling little volume, keeping the mass-weighted $\tau_{\rm cool}/\tau_{\rm ff}$ low even after a stable hot phase is already established. In the Feedback case, the hot phase is predominantly at $\tau_{\rm cool}/\tau_{\rm ff}>1$ already at high redshift and evolves only weakly thereafter, so the subsequent rise of the all-gas mass-weighted ratio is driven largely by the declining mass contribution of the cold phase. This distinction therefore helps disentangle the evolution of the volume-filling hot corona from changes in the relative mass contribution of the cold component.
  
In Section~\ref{sec:outside-inVIR} we demonstrated that when the evolution of the CGM is studied using radii scaled by $R_\text{vir}$, the Feedback run shows an outside-in virialisation pattern, i.e. the outer CGM reaches $\tau_\text{cool}/\tau_\text{ff}\!>\!1$ before the inner CGM, see Fig.~\ref{FIG:SCALES}.  Thus, when feedback is included, the {\small VINTERGATAN} suite behaves consistently with the MW-type haloes from the {\small FIRE} simulations \citep{Stern2021,Sultan2024arXiv241016359S}. The No-FB runs with fixed metallicities show the opposite behavior: virialisation proceeds inside-out, recovering the classical picture of accretion-shock heating from \citetalias{Dekel2006}. This suggests that feedback inverts the radial sequence of CGM virialisation, a point that has not been clearly emphasized in the literature. Nevertheless, the role of feedback in this process has been discussed in the review by \citealt{Faucher2023ARA&A..61..131F}, based on results from idealized analytical modeling and the \textsc{FIRE} simulations.

However, as shown in Fig.~\ref{FIG:SCALES}, the evolution of $\tau_{\text{cool}}/\tau_{\text{ff}}$ changes markedly depending on whether the CGM is measured using shells scaled by $R_{\text{vir}}$ or shells in fixed physical kiloparsecs (proper scale). In practice, the CGM is largely established already by $z\!\sim\!3$ far beyond the virial radius at that epoch, with later evolution occurring mainly at radii of $\sim\!30$--$50$ kpc, where the cold inner structure gradually evolves. This qualitative picture also holds for different halo assembly histories, although Fig.~\ref{FIG:Vir_EarlyLate} shows that the transition can range from nearly halo-wide to rapid outside-in depending on the run. Thus, a general picture of gradual outside-in virialisation is not always directly applicable in our runs. Instead, feedback-driven virialisation is better described as a rapid transition across most of the halo. Depending on the run, this transition can appear nearly halo-wide or show a short outside-in sequence, but in either case it is largely completed within $\lesssim 0.5$ Gyr.

The reason for this discrepancy comes down to $R_{\text{vir}}$ being defined as an evolving quantity\footnote{$R_{\rm vir}= \left(\frac{3M_{\rm vir}}{4\pi \Delta_{\rm vir}\rho_c(z)}\right)^{1/3}$ with the critical radius $\rho_c(z) = \frac{3H(z)^2}{8\pi G}\propto (1+z)^3$, where the latter proportionality hold during the matter dominated epoch ($z\sim 0.5-1000$).}, regardless of any actual underlying physical evolution of the dark matter halo. In Fig.~\ref{FIG:HISTOGRAMS},  $R_{\text{vir}}$ (dashed white line), $0.2R_{\text{vir}}$ (solid white line), and the dark matter halo scale radius $r_{\rm s}$ (purple line) are overplotted on the histograms. As noted above, the underlying thermal structure of the halo is established early, with the transition between cold and hot gas shifting only by 10-20 kpc over time. However, when this slowly evolving structure is examined using an evolving scale such as $R_{\text{vir}}$, it creates the misleading impression that the CGM is evolving more strongly than it actually is. In the Feedback run, this is particularly evident around $z\!=\!0.5$, when the $0.2R_{\text{vir}}$ shell moves from tracing the cold inner phase to tracing the hot, virialized CGM, even though the cold inner phase itself has not changed nearly as much in physical extent. We refer to this effect as \emph{``CGM pseudo-evolution''}.

This concept builds on earlier studies of pseudo-evolution of dark matter haloes \citep{Diemer2013ApJ...779..159D,Diemer_2_2013}, an effect that affects conclusions drawn on mass growth across all halo masses, but more crucial for low-mass systems that assemble early. The authors demonstrated that for dark matter haloes with masses of $\sim10^{12}\, {\rm M}_\odot$, relevant for our work, dark matter density profiles do not physically evolve much within many 100s of kiloparsecs below $z\sim 1$; their underlying structure is set at early times, and any late-time "growth" is pseudo-evolution due to $R_{\text{vir}}$, and therefore $M_{\text{vir}}$, evolving. Connecting to our simulations, while $R_{\text{vir}}$ grows with time, the dark matter scale radius is not evolving for $z\lesssim 3$ \citep[see also][]{Wang2024MNRAS.52710760W}, indicating that the inner structure of the halo is established at this epoch. We find that the onset of the virialized hot corona closely coincides with this stabilization: in all simulations, $r_{\rm s}$ becomes nearly constant at $\sim\!20$--$25\ \text{kpc}$ by $z\!\sim\!3$. Whether this temporal correspondence reflects a causal connection is unclear. Both processes may instead be linked to the halo assembly history and the associated evolution of the gravitational potential.
  
Together, these results suggest that much of the apparent late-time CGM evolution inferred from $R_{\text{vir}}$-based shells is driven by the continued growth of $R_{\text{vir}}$, rather than by large changes in the physical thermodynamic state of the CGM. Therefore, fixed physical apertures provide a complementary view of the CGM evolution and show that the hot corona is largely established earlier than implied by $R_{\text{vir}}$-scaled measurements. To our knowledge, this is the first explicit identification of pseudo-evolution affecting the characterization of the thermodynamic state of the CGM. This complements recent work showing the importance of pseudo-evolution for the evolution of the enclosed gas fraction in more massive halos \citep{Pontzen2026arXiv260516483P}.

\subsection{ Caveats}
The {\small VINTERGATAN} cosmological simulation suite has several caveats that add uncertainties on characterizing the onset of the hot corona and the virialisation of the CGM. 

Our simulations neglect several physical agents. Active Galactic Nuclei (AGN) feedback can affect CGM virialisation through gas heating and entropy injection. AGN can enhance the ejection of hot gas into the IGM, sustain a hot medium, and quench star formation. Its impact depends on the ratio between kinetic and thermal energy injection into the CGM \citep{Prasad2026MNRAS.545f2155P} and on the halo assembly history \citep{Davies2021MNRAS.501..236D}. However, for the halo masses considered here at the time of CGM virialisation ($<10^{12}\,{\rm M_\odot}$), AGN feedback is not expected to significantly affect the bimodal distribution of $\tau_\text{cool}$ in the CGM. This bimodality is associated with the coexistence of hot and cold accretion modes, and is likely driven mainly by stellar feedback while the central black hole grows only modestly \citep{Correa2018MNRAS.473..538C,Wright2024}. 

Cosmic rays (CR) can influence the CGM by building pressure gradients that enhance pressure support of cold inflows and drive the ejection of cold gas from the ISM into the CGM \citep{Everett2008ApJ...674..258E,Booth2013ApJ...777L..16B, Naab2017ARA&A..55...59N, Bieri2026MNRAS.547ag216B}. They can significantly reduce temperatures in the inner CGM ($<150$ kpc), shortening $\tau_\text{cool}$ and affecting inner virialisation \citep{Ji2021MNRAS.505..259J, Ruszkowski2023A&ARv..31....4R}. Their importance depends on redshift and halo growth. For Milky Way–mass haloes at the epochs when the hot corona forms (in this study), CR pressure is generally insufficient to balance gravity \citep[CR are negligible at $z\!\gtrsim\!1\!-\!2$]{Hopkins_c2020MNRAS.492.3465H, Ji2020MNRAS.496.4221J}. However, CR become relevant at later times within $\sim300$ kpc \citep{Ji2021MNRAS.505..259J}. Therefore, our simulations may differ in inner CGM properties at low redshift if CR were included, although predictions remain uncertain due to poorly understood CR transport physics \citep{Hopkins2026OJAp....961465H}.

Magnetic fields can also influence CGM dynamics, for example by slowing gas flows and increasing the density of the inner CGM \citep{Voort2021MNRAS.501.4888V}, thereby modifying the interaction between cold and hot gas phases \citep{Faucher2023ARA&A..61..131F}. This means that it affects mass growth, cold-cloud velocities, and cloud survival. For typical CGM field strengths, aligned fields have little impact on cold clouds, whereas transverse fields can significantly slow them and help the cold phase persist longer with high densities \citep{Kaul2025MNRAS.539.3669K}.

Numerical resolution is another intrinsic limitation. If the CGM is not sufficiently resolved, cold gas structures may be artificially suppressed, and potentially leading to discrepancies with observational surveys like COS-LRG or SDSS \citep{Chen2018MNRAS.479.2547C,Zahedy2019MNRAS.484.2257Z, Lan_2014, Lan_2018} by suppressing the simulated covering fraction of cold-gas tracers such as Mg~II absorbers \citep{Nelson2020MNRAS.498.2391N}. Insufficient resolution also prevents a full capture of cold–hot phase interactions \citep{Mandelker2021gcf2.confE..63M} and the turbulent structure of the cold gas \citep{Gronke2020MNRAS.492.1970G}. Although insufficient resolution can suppress cold gas structures and phase mixing, the hot CGM is a volume-filling phase and should be less strongly affected in its bulk thermodynamic properties \citep{Rey2024, Singh2024MNRAS.532.3222S}.

Finally, our results on the early onset of the hot corona and CGM virialisation depend on how stellar feedback is implemented. Variations in feedback and hydrodynamics can alter, for example, the CGM baryon and metal content \citep[][ within the AGORA project \citealt{Kim2014ApJS..210...14K, Roca2021ApJ...917...64R}]{StrawnAGORA2024ApJ...962...29S}, while different energy injections can modify its enrichment history \citep[][using gas-regulator model arguments]{Carr_2023}, and thus the cooling properties. Whether our results are robust to different feedback prescriptions will be tested in future comparative work within the AGORA framework.


\section{Conclusions}\label{sec:summary}
In this work we have used cosmological simulations of Milky Way-mass galaxies from the {\small VINTERGATAN}  suite to study formation of the circumgalactic medium (CGM), with a focus on the roles played by stellar feedback processes and accretion shocks in establishing the hot corona and CGM virialisation. Simulations were carried out both with and without feedback and compared with the analytical Virial Shock model of \citetalias{Birnboim2003MNRAS.345..349B,Dekel2006}.
Our main results are summarized as follows:

\begin{enumerate}
\item Stellar feedback triggers hot-corona formation much earlier than classical accretion-shock models predict. With feedback, the CGM reaches $T\gtrsim T_{\rm vir}$ across several hundred kpc by $z\!\sim\!3$--$4$, whereas the Virial Shock model predicts this transition only at $z\!\sim\!1$ for $Z_{\rm CGM}=0.1Z_\odot$. Without feedback, the hot corona forms later and more gradually, with the volume-filling gas at $R_{\rm vir}$ cooler by $\sim\!1$ dex at $z\!\sim\!4$. The NoFB runs show a strong metallicity dependence and broadly follow the trends expected from \citetalias{Birnboim2003MNRAS.345..349B,Dekel2006}, although the analytical model overpredicts the halo mass required for hot gas to reach $R_{\rm vir}$.

\item Feedback and feedback-free runs show similar hot-corona temperatures within $R_{\rm vir}$ at $z=0$, when the haloes reach Milky Way-like masses. This apparent agreement with the \citetalias{Birnboim2003MNRAS.345..349B,Dekel2006} prediction hides different heating histories, so the present-day CGM alone cannot reveal which heating mechanism was dominant. Following the full evolution shows that early stellar feedback shifts the transition from a cold CGM to a hot corona toward lower halo masses, while feedback-free runs reach this state later and at higher masses.

\item CGM virialisation, defined by $\tau_{\rm cool}/\tau_{\rm ff}>1$, closely tracks hot-corona formation, but not necessarily the direction of shock propagation. In accretion shock-dominated runs, virialisation proceeds inside-out following the propagation of the shock. Stellar feedback alters this behavior, driving virialisation over $\lesssim 0.5$ Gyr. During this rapid transition, virialisation can occur nearly simultaneously across most of the CGM or follow an outside-in sequence, with $\tau_{\rm cool}/\tau_{\rm ff}>1$ reached first at large radii and later in the inner CGM. Our results are in agreement with previews studies on CGM virialization as \citealt{Stern2021, Sultan2024arXiv241016359S}.

\item The inferred virialisation sequence depends on how the CGM is defined. In $R_{\rm vir}$-scaled shells, the Feedback run takes $\sim5$ Gyr to fulfill $\tau_{\rm cool}/\tau_{\rm ff}>1$ within $R_{\rm vir}$, while the NoFB runs appear to virialize almost simultaneously within $0.5R_{\rm vir}$. In fixed physical shells, the Feedback run instead shows a rapid transition within 300 kpc, whereas the NoFB runs retain a clear inside-out pattern with strong metallicity dependence. This transition coincides with the stabilization of the dark matter halo scale radius, suggesting that the inner halo and CGM are largely established by $z\!\sim\!3$.

\item The contrast between $R_{\rm vir}$-scaled and fixed-kpc shells reveals a pseudo-evolution of the CGM. After the hot corona is established, $R_{\rm vir}$ continues to grow even though the gas properties within 300 kpc evolve only weakly. As a result, fixed fractions such as $0.2R_{\rm vir}$ can trace the evolving halo boundary rather than genuine gas evolution. This is analogous to pseudo-evolution in dark matter haloes \citep{Diemer2013ApJ...779..159D,Diemer_2_2013}; to our knowledge, its impact on CGM virialisation diagnostics has not been explicitly reported before.
\end{enumerate}

Future work will explore how early CGM virialisation shapes the later hot corona, including its density structure, temperature distribution, metal mixing, and gas kinematics. We will also connect early hot-corona formation with the baryon cycle, the evolution of cold gas in galaxies, and galactic gas disc emergence at high redshift, building on previous work linking inner virialization to the formation of thin galactic disc \citep{Stern2021, Hafen22}.

\section*{Acknowledgements}
VRP, OA and SRF acknowledge support from the Knut and Alice Wallenberg Foundation, the Swedish Research Council (grant 2025-04892), the Swedish National Space Agency (SNSA grants 2025-00181, 2023-00164 and 2025-00405), the LMK foundation, eSSENCE, a Swedish strategic research program in e-Science, and the Walter Gyllenberg Foundation's fund (grants 2024-0026, 2024-0035 and 2025-46414). 

We thank Jonathan Stern, Romain Teyssier, Filippo Fraternali, Tommaso Zana, Mahsa Kohandel and Andrea Ferrara for insightful discussions and comments that helped improve this work. We also thank the referee for their constructive comments and suggestions, which helped improve the clarity and interpretation of our results.

\section*{Data availability}
The IllustrisTNG data used in this article are available at https://www.tng-project.org, and we will share derived data products upon reasonable requests. We thank Martin Rey for providing these data, shown in \cite{Martin2022MNRAS.510.4208R}.

\bibliographystyle{mnras}
\bibliography{article_ref} 



\appendix

\section{Metallicity dependence of hot corona expansion}
\begin{figure*}
    \centering
    \includegraphics[width=1\textwidth]{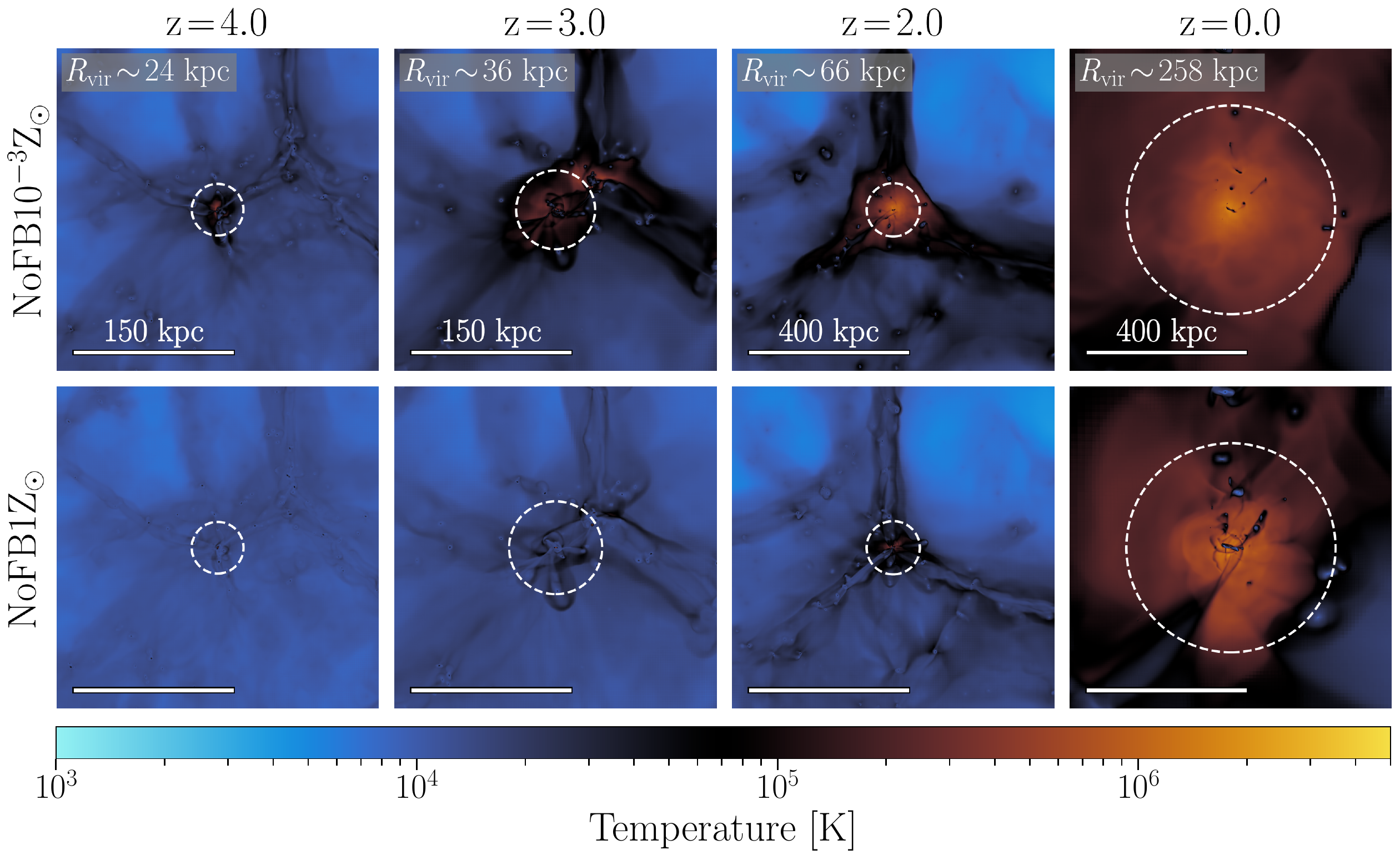}
    \caption{Same as Fig.~\ref{FIG:MULTITEMP}, but for the NoFB$10^{-3}Z_\odot$ and NoFB$1Z_\odot$ runs.}
    \label{FIG:MultuTemp_NoFBrest}
\end{figure*}
In this appendix, we complement Fig.~\ref{FIG:MULTITEMP} with Fig.~\ref{FIG:MultuTemp_NoFBrest}, showing density projections colored by temperature for the NoFB$10^{-3}Z_\odot$ and NoFB$1Z_\odot$ runs. These projections illustrate how metallicity regulates hot-corona formation when accretion shocks are the only heating mechanism. Increasing the metallicity leads to a progressively delayed and more compact hot phase. At $z=4$, the very low-metallicity run, NoFB$10^{-3}Z_\odot$, already contains extended warm-hot gas within $R_{\rm vir}$, whereas NoFB$1Z_\odot$ shows no visible hot phase. By $z=3$, NoFB$10^{-3}Z_\odot$ has filled the volume enclosed by $R_{\rm vir}$ with warm-hot gas, reaching temperatures of $\sim 10^{5}$--$10^{6}$ K. We also identify extended hot gas entering $R_{\rm vir}$ through an ongoing merger. In contrast, the NoFB$1Z_\odot$ run retains predominantly cold gas within $R_{\rm vir}$. By $z=2$, NoFB$1Z_\odot$ has still not formed a hot corona extending to, or slightly beyond, $R_{\rm vir}$; hot gas is only beginning to develop around the halo center. By $z=0$, however, a hot corona filling the CGM is established in both cases. The presence of hot gas far beyond $R_{\rm vir}$ is noteworthy because Fig.~\ref{FIG:MSHOCK_METALS} does not predict a stable shock at $R_{\rm vir}$ for these epochs. This reinforces the limitation of the Virial Shock model suggested in Sec.~\ref{sec:expectedOnset}.

\section{Dependency of CGM gas virialisation on halo assembly history}
\label{sec:ap_AMH}
\begin{figure*}
    \centering
    \includegraphics[width=0.9\textwidth]{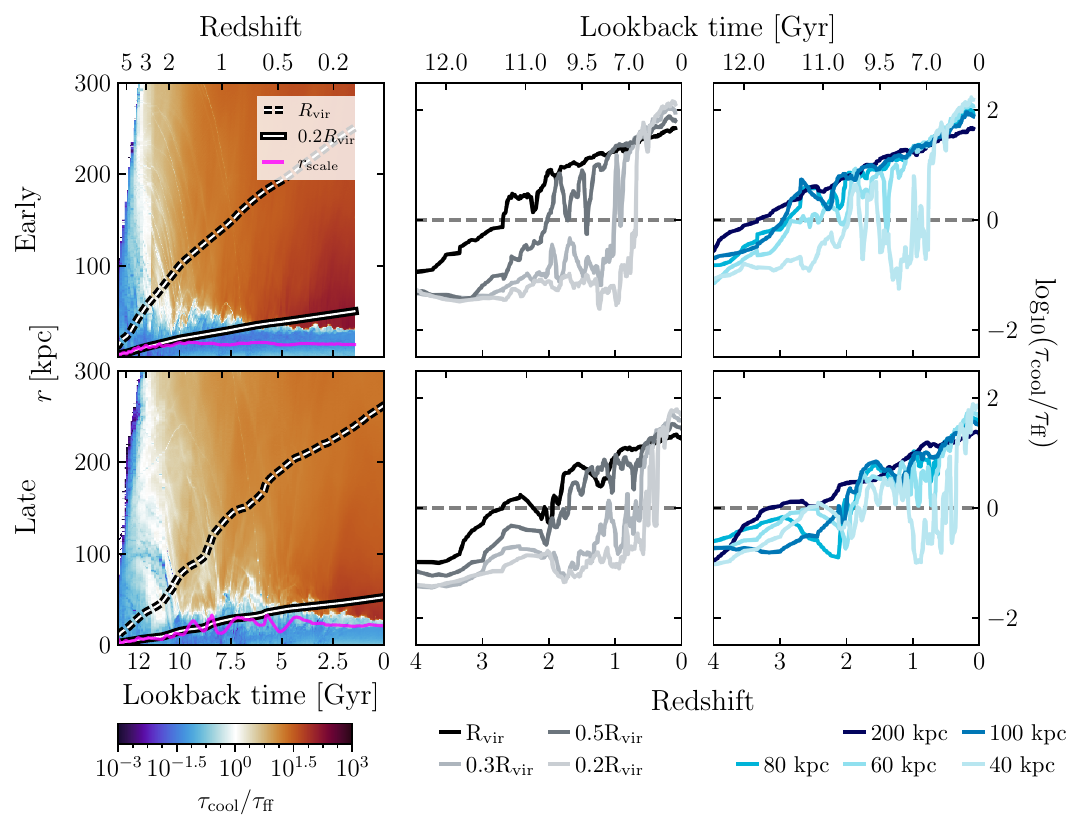}
    \caption{Same as Fig.\ref{FIG:VIRIALIZATION_HIST} and Fig.~\ref{FIG:SCALES}, but for the Early and Late halo assembly histories. }
    \label{FIG:Vir_EarlyLate}
\end{figure*}

In this study, an important aspect of the analysis is to quantify how the halo mass assembly history affects the onset of the hot corona and the virialisation process of the CGM and halo gas. Figure~\ref{FIG:Vir_EarlyLate} shows, in the first column, the same analysis of $\tau_{\rm cool}/\tau_{\rm ff}$ as in Fig.~\ref{FIG:VIRIALIZATION_HIST}, but for the Early and Late runs. virialisation proceeds in a broadly similar way in both cases, indicating that the overall transition does not depend strongly on differences in halo mass assembly history. The remaining columns repeat the analysis of Fig.~\ref{FIG:SCALES} for the Early and Late runs. Overall, both runs follow the same qualitative behavior as the Feedback run, but with differences in the timing of the transition. The Early run reaches a virialized state at $R_{\rm vir}$ earlier than the Late run and appears more consistent with an outside-in sequence. By contrast, the Late run exhibits a more rapid, nearly halo-wide virialisation, with several fixed physical shells crossing $\tau_{\rm cool}/\tau_{\rm ff}>1$ over a relatively short time interval.

This figure strengthens the results presented in Sec.~\ref{sec:outside-inVIR}. It confirms the outside-in virialisation pattern in $R_{\rm vir}$-based shells, while the fixed physical shells show that feedback-driven virialisation can occur rapidly across a large fraction of the halo. The comparison between the Early and Late runs indicates that halo assembly history mainly shifts the timing of the transition, without changing the overall feedback-driven virialisation scenario. The use of $R_{\rm vir}$-based shells therefore continues to highlight the impact of CGM pseudo-evolution.

\section{The enrichment of the CGM in {\small VINTERGATAN}}\label{SEC:AP_METALS}
\begin{figure*}
    \centering
    \includegraphics[width=0.9\textwidth]{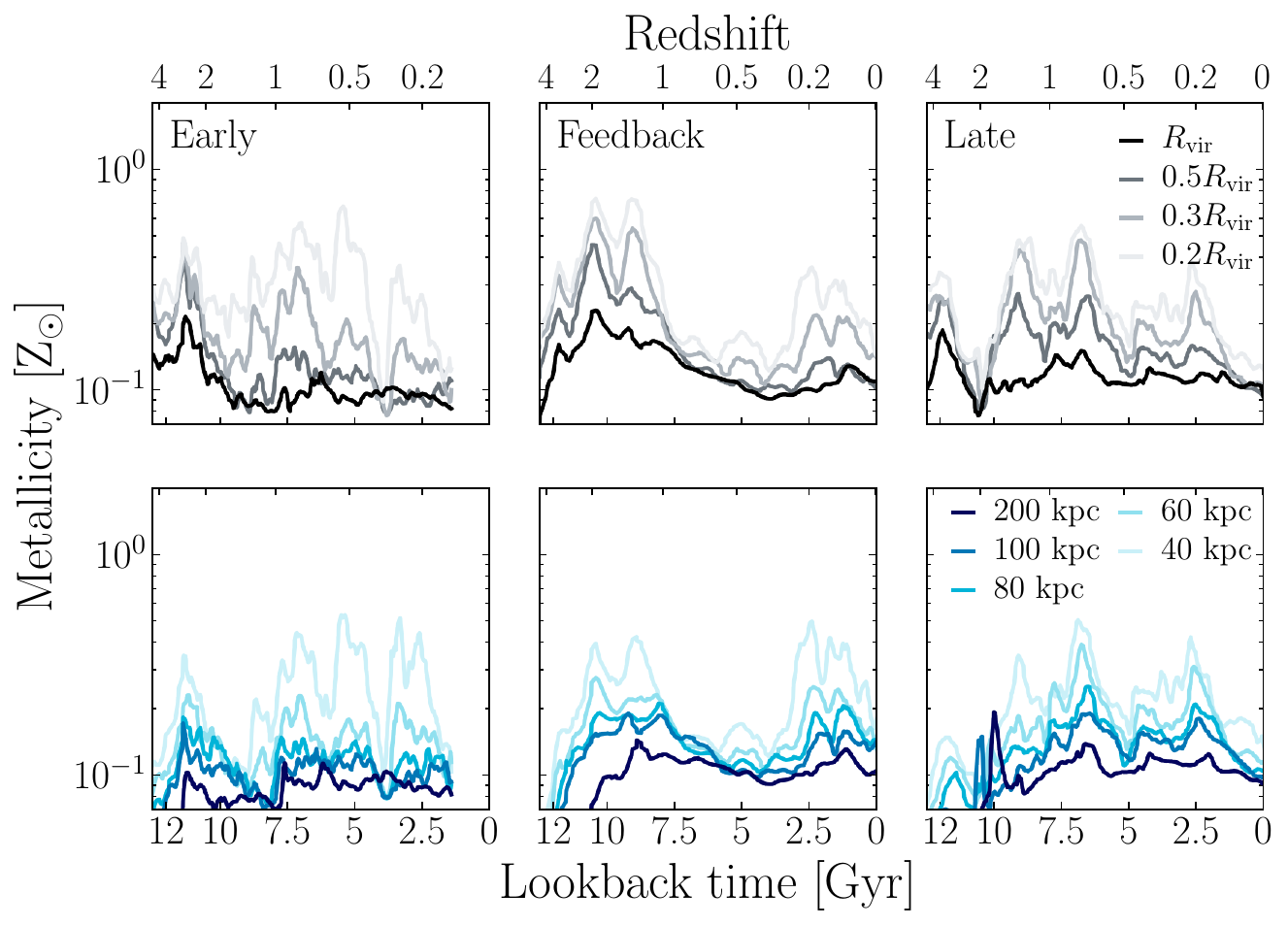}
    \caption{Time evolution of the average metallicity in $R_\text{vir}$-based shells (top panels) and fixed physical (proper) shells (bottom panels) for feedback runs.  }
    \label{FIG:Metallicity_evolving}
\end{figure*}

Here we examine how enrichment evolves in the Early, Feedback, and Late runs, since metallicity strongly affects gas cooling. Fig.~\ref{FIG:Metallicity_evolving} shows the average metallicity in the same shells as Fig.~\ref{FIG:SCALES}: the top panels use $R_{\rm vir}$-based shells, while the bottom panels use fixed physical radii. In the Feedback run, which shares the same halo formation history as the NoFB runs described in Sec.~\ref{sec:method_sim}, the representative metallicity between 60 and 200 kpc is roughly $0.1Z_\odot$ over much of the evolution. This motivates the NoFB$0.1Z_\odot$ run as the closest comparison case for isolating the effect of stellar feedback. At large radii, the three feedback runs reach similar metallicities despite different enrichment histories, suggesting that CGM metallicity differences are not the main driver of their different virialisation times.

\section{Comparison to cooling flow models} \label{sec:ap_coolingflow}
Complementary to the Virial Shock model, we use the Cooling Flow model following \citet[][hereafter \citetalias{Stern2019,Stern2020}]{Stern2019,Stern2020}, as a benchmark for the stability of an already existing hot corona. The model defines a critical hot accretion rate, $\dot{M}_{\rm crit}$, and an associated threshold halo mass, $M_{\rm thres}$, below which radiative cooling removes thermal pressure support faster than the corona can maintain hydrostatic balance. We estimate $\dot{M}_{\rm crit}$ and then $M_{\rm thres}$ using halo and CGM gas properties measured directly from our simulations. In our study, we do not attempt to fit the full cooling-flow solution of each run. Instead, we use the model as a benchmark prediction for how $M_{\rm thres}$ depends on redshift, metallicity, and $f_{\rm CGM}$. We evaluate the relevant CGM quantities at 60 kpc\footnote{In \citetalias[][and \cite{Stern2021}]{Stern2020}, $R_{\mathrm{circ}}=20$ kpc, corresponding roughly to $\sim0.1R_{\rm vir}(z=0)$. In our simulations, this radius often overlaps with the ISM.}.

To obtain $M_{\rm thres}$, we follow sec.~3 of \citetalias{Stern2020} and define the threshold mass as the halo mass for which
\begin{equation}
\label{eq:mdotcrit}
    \dot{M}_{\mathrm{crit}}(M_\text{thres},z) =
    f_\text{CGM}\, f_\text{bar}\, \frac{M_\text{vir}}{t_\text{H}},
\end{equation}
where $f_\text{CGM}$ is the CGM mass fraction, $f_\text{bar}$ is the cosmic baryon fraction, and $t_\text{H}$ is the Hubble time. The right-hand side gives a characteristic baryonic accretion rate associated with the hot CGM reservoir. At each epoch, we vary $M_\text{vir}$ while fixing $f_\text{CGM}$ and the hot-phase metallicity to the values measured in the simulations, and keep the remaining parameters as in \citetalias{Stern2020}. 

To solve Eq. \ref{eq:mdotcrit}, we evaluate $\dot{M}_{\mathrm{crit}}$ as shown below in \ref{eq:CoolingFlow}, finding for which value of $M_\text{vir}$, binned from $10^{10}$--$10^{13} {\rm M_\odot}$, this condition is satisfied. We then identify this $M_\text{vir}$ as $M_\text{vir}\equiv M_{\rm thres}$. For consistency with the `benchmark prediction' shown in Fig.~9 of \citetalias{Stern2020}, we compute the critical accretion rate using their approximated definition for virial velocity and $R_{\rm vir}$ \citepalias[eqs. 31 and 32 of][]{Stern2020},
\begin{equation}
    \dot{M}_{\mathrm{crit}} \approx 
    10.6 \, f_{\mathrm{v_c}}^3 \, f_\lambda \lambda_{0.035} \,
    \Lambda_{-22}^{-1}\, M_{12}^{5/3} \, E(z)^{2/3} \,
    \mathrm{M}_{\odot} \, \mathrm{yr}^{-1},
\label{eq:CoolingFlow}
\end{equation}
where $M_{12}=M_{\rm vir}/10^{12}\,{\rm M_\odot}$, 
$f_{\mathrm{v_c}} =v_\text{c}(R_{\text{circ}})/v_\text{vir}$, 
$f_\lambda$ accounts for differences between the specific angular momentum of the baryons and the halo average, 
$\lambda_{0.035}$ is the halo spin parameter normalized to 0.035, 
$\Lambda_{-22}=\Lambda/10^{-22}\,{\rm erg\,cm^3\,s^{-1}}$, 
and $E(z)$ is the cosmological evolution factor. We evaluate $\Lambda$ using the cooling tables implemented in {\small RAMSES}, instead of those of \cite{Wiersma2009MNRAS.393...99W} adopted by \citetalias{Stern2020}, so that the model uses the same thermodynamics as the simulations.

\begin{figure*}
    \centering
    \includegraphics[width=\textwidth]{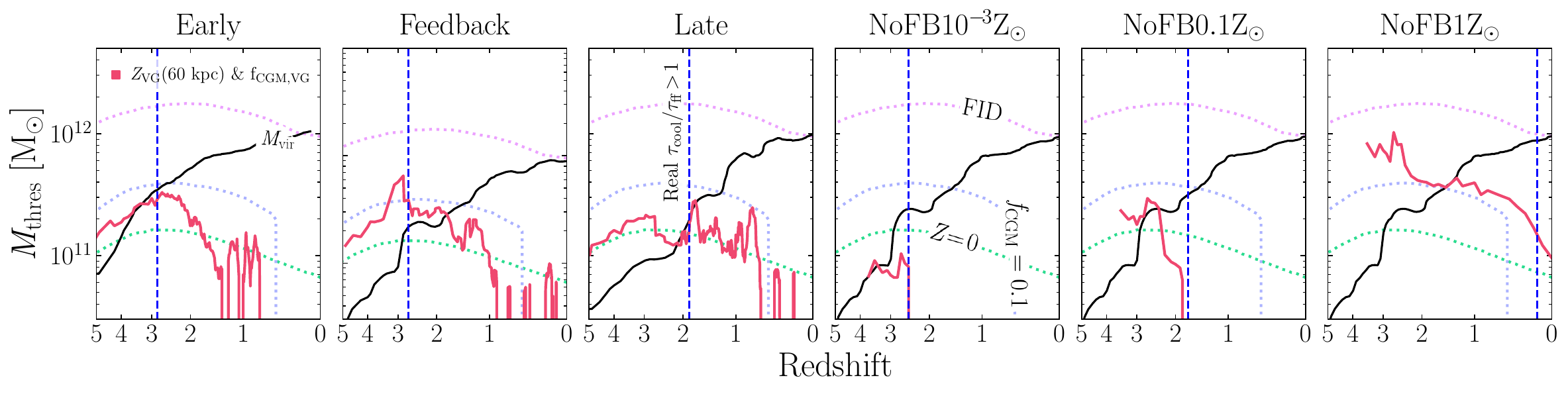}
\caption{The panels show the expected $M_{\rm thres}$, computed using the metallicity and $f_{\rm CGM,VG}$ measured in the simulations while adopting the parameter definitions of \protect\citetalias{Stern2020}. This estimate is shown as a red line. The halo mass evolution, $M_{\rm vir}$, is shown as a black solid line. The blue dashed line marks when $\tau_{\rm cool}/\tau_{\rm ff}\!>\!1$ at 60 kpc for our simulations, see Fig. \ref{FIG:SCALES}. For comparison with \protect\citetalias{Stern2020}, the faint dotted curves show our calculation of $M_{\rm thres}$ for three analytic cases from Fig.~9 by \protect\citetalias{Stern2020}: $f_{\rm CGM}=1$ (fiducial, purple), no metals ($Z\!=\!0$, green), and $f_{\rm CGM}=0.1$ (blue). The metallicity for the dotted line cases is computed following eq. 37 in \protect\citetalias{Stern2020} based on the mass–metallicity relation from \protect\cite{Andrews2013ApJ...765..140A}, and the stellar-mass–halo-mass relation from \protect\cite{Behroozi19}. } 
    \label{FIG:COOL_FLOW_test}
\end{figure*}

In Fig.~\ref{FIG:COOL_FLOW_test}, we compare $M_{\rm thres}$ with the virial-mass growth, $M_{\rm vir}$, of our runs. We also include three reference scenarios from \citetalias{Stern2020}: $f_{\rm CGM}=1$, labeled as fiducial (FID), $Z=0Z_\odot$, and $f_{\rm CGM}=0.1$. The vertical dashed line marks the epoch at which each simulation first satisfies $\tau_{\rm cool}/\tau_{\rm ff}>1$ at 60 kpc, a close value, as measured in Fig.~\ref{FIG:SCALES}.

We first focus on when the intersection between $M_{\rm vir}$ and $M_{\rm thres}$ occurs. For the feedback runs, this crossing occurs at $3\gtrsim z\gtrsim2$, broadly close to the measured $\tau_{\rm cool}/\tau_{\rm ff}>1$ transition. The agreement is not exact. In the Early and Late runs, the predicted crossing occurs close to the measured transition, whereas in the Feedback run the model predicts stabilization later than in the simulation. This suggests that halo mass alone, here connected to different halo-formation histories, is not necessarily the only parameter controlling the onset of hot-corona stability in individual haloes.

The Late run also highlights a degeneracy in the model interpretation. Despite having higher metallicity than the $Z=0Z_\odot$ reference case (Fig.~\ref{FIG:Metallicity_evolving}), its predicted $M_{\rm thres}$ follows closely the analytic no-metallicity case with $f_{\rm CGM}=1$. This indicates that different combinations of metallicity and CGM gas fraction can yield similar values of $M_{\rm thres}$, making it difficult to isolate which physical ingredient primarily drives the stabilization. A more detailed analysis of the roles of metallicity and $f_{\rm CGM}$ will be presented in Rufo-Pastor et al. (in prep.).

For the NoFB runs, the predicted crossing shows poorer agreement with the measured transitions. In these cases, changes in metallicity produce a clear shift in $M_{\rm thres}$, but the model does not reliably capture the simulated stabilization times. The strongest discrepancy occurs for NoFB$1Z_\odot$, where the model predicts an earlier transition, at $z\approx1.5$, than the one indicated by the simulation, at $z<0.5$. In addition, the measured transition occurs near values representative of the fiducial analytic case with $f_{\rm CGM}=1$ and $Z\approx Z_\odot$, while for NoFB$0.1Z_\odot$ it shifts toward the analytic $f_{\rm CGM}=0.1$ case at the same metallicity.

Overall, the Cooling Flow framework broadly recovers the early transition in the feedback runs, but performs poorly for the NoFB cases. This suggests that the Cooling Flow threshold is most useful as a qualitative benchmark when feedback maintains a multiphase CGM, where a pressure-supported hot atmosphere coexists with a colder inner phase. The predicted virialisation sequence therefore appears to depend on the survival of the inner cold phase, which is regulated by how stellar feedback transports metals and energy into the CGM. This dependence will be explored further in comparative work by Lindh et al. (in prep.).

\bsp	
\label{lastpage}
\end{document}